\documentclass[3p]{elsarticle}

\usepackage{lineno,hyperref}
\usepackage{amsmath}
\usepackage{comment}
\usepackage{float}
\newtheorem{theorem}{Theorem}
\usepackage{subfigure}

\modulolinenumbers[50]

\journal{Journal of \LaTeX\ Templates}

\biboptions{authoryear}

\begin{document}

\begin{frontmatter}

\title{On The Limitation of Some Fully Observable Multiple Session Resilient Shoulder Surfing Defense Mechanisms
 \tnoteref{mytitlenote}}

\author[add1]{Nilesh Chakraborty\corref{cor1}}
  \ead{nilesh.pcs13@iitp.ac.in}
\author{Samrat Mondal}
  
  \cortext[cor1]{Corresponding author}
  \address{Department of Computer Science and Engineering}
  \address{Indian Institute of Technology, Patna}
  \address{Bihar - 801103, India}

\begin{abstract}
Using password based authentication technique, a system maintains the login credentials (username, password) of the users in a password file. Once the password file is compromised, an adversary obtains both the login credentials. With the advancement of technology, even if a password is maintained in hashed format, then also the adversary can invert the hashed password to get the original one. To mitigate this threat, most of the systems now a days store some system generated fake passwords (also known as \textit{honeywords}) along with the original password of a user. This type of setup confuses an adversary while selecting the original password. If the adversary chooses any of these \textit{honeywords} and submits that as a login credential, then system detects the attack.

A large number of significant work have been done on designing methodologies (identified as $\text{M}^{\text{DS}}_{\text{OA}}$) that can protect password against observation or, shoulder surfing attack. Under this attack scenario, an adversary observes (or records) the login information entered by a user and later uses those credentials to impersonate the genuine user. In this paper, we have shown that because of their design principle, a large subset of $\text{M}^{\text{DS}}_{\text{OA}}$ (identified as $\text{M}^{\text{FODS}}_{\text{SOA}}$) cannot afford to store \textit{honeywords} in password file. Thus these methods, belonging to $\text{M}^{\text{FODS}}_{\text{SOA}}$, are unable to provide any kind of security once password file gets compromised. Through our contribution in this paper, by still using the concept of \textit{honeywords}, we have proposed few generic principles to mask the original password of $\text{M}^{\text{FODS}}_{\text{SOA}}$ category methods . We also consider few well established methods like S3PAS, CHC, PAS and COP belonging to $\text{M}^{\text{FODS}}_{\text{SOA}}$, to show that proposed idea is implementable in practice.

\end{abstract}


\end{frontmatter}

\linenumbers

\section{Introduction}
\label{sec:introduction}
Password based authentication remains as one of the most dominant forms of identity verification since last few decades \citep{Upassword}. However, this scheme is found to be vulnerable under different types of attack \citep{pguess} \citep{phishing} \citep{key-logger} \citep{dictionary-attack}. Inversion attack is one such lately developed attack model on the password based schemes which threats its security standard to a great extent \citep{weir} \citep{compromised-password-file-e3}. \\

\textbf{Inversion attack:} 
Once the password file (identified as $F_P$: on Unix systems the file $F_P$ might be /etc/passwd or /etc/shadow) gets compromised, an adversary may found that the passwords are either maintained in plaintext or, in the hashed format. If the passwords are stored in plaintext, then the adversary can directly impersonate all the users.
Even if the passwords are maintained in hashed format, then also the adversary may conduct guessing attack to invert (i.e., deriving P from H(P): where H(P) denotes a hashed password) those. Initially, to break a password, brute force attack was conducted by guessing many possible combinations. But the time complexity of brute force search technique used to be very high as an attacker tried for every possible option to crack a password. In $2008$, proposed algorithm by John the Ripper significantly reduced the complexity of password guessing process compared to the brute force search technique \citep{john-password-cracking}. In $2009$, using the concept of probabilistic context free grammar, Weir et al. were able to crack $28\%-129\%$ more passwords than John the Ripper password cracking algorithm \citep{weir}. Recently, based on Markov chain model, modern password crackers further improve the inversion rate \citep{ma}. \\


\textbf{Evidence:} There are some strong evidence of inversion attack in the recent past. In $2013$, almost $50$ million passwords of Evernote were compromised under this attack model \citep{evernote}. Giant web based organizations like LinkedIn, Yahoo, RockYou have gone through the same misery \citep{compromised-password-file-e3}. Thus, there was urgency in developing an authentication technique which can address this attack. Till date, \textit{Honeyword Based Authentication Technique} (\textbf{HBAT}) is the best known approach which mitigates this threat to a great extent \citep{honeyword-juels}.\\

\textbf{HBAT framework:} In HBAT framework, along with the original password, a system maintains another $k-1 (k \geq 2)$   dummy passwords for each user's account. These system generated dummy passwords are also known as \textit{honeywords}. The original password and the \textit{honeywords} are collectively known as \textbf{\textit{sweetwords}}. All these \textit{sweetwords} are very close to each other, but they are not exactly the same. Let the username and password of a user be \textit{alex} and \textit{alex1992}, respectively. Then the system may maintain the following list of \textit{k} (considered as $6$ here) \textit{sweetwords} against the username \textit{alex}.

\begin{center}
\begin{tabular}{c c c}
alex1990 & alex1994 & alex1991\\
\textbf{alex1992} & alex1995 & alex1993
\end{tabular}
\end{center}

The index of the original password (here $4$, starting from $1$), along with username (\textit{alex}), is maintained in another file in a different server, known as \textit{honeyChecker}. 

Therefore, under the HBAT framework, an adversary gets confused among \textit{k} ($>1$) \textit{sweetwords} by looking at a compromised $F_P$. In order to login successfully, an adversary needs to guess the original password correctly from the list of $k$ probable options. This yields probability of selecting the correct password as $1/k$.\\

\textbf{HBAT threat detection strategy:} Let us assume that the guessed  and the submitted password by the adversary belongs to the $t^{th}$ ($1 \leq t \leq k$) index position in the list of $k$ \textit{sweetwords}. From the  password submitted by the adversary, system then retrieves the index position (here $t$) of the password. Along with the username, the index value is then communicated to the \textit{honeyChecker}.  For the targeted username, if the received index value gets matched with the stored one, the \textit{honeyChecker} then  sends a positive feedback signal to the system administrator otherwise, sets off a security alarm by detecting the security breach.\\

In a nutshell, by maintaining a list of $k$ \textit{sweetwords} against each user's account, the success probability of an attacker to crack the original password  is reduced to $1/k$. As HBAT maintains the password information in two different servers, thus it provides a distributed security framework which is harder to compromise as a whole \citep{sauth} \citep{honeyword-juels}.\\

\textbf{Motivation and Contribution:} 
Most of the recent researches indicate that along with the original passwords, there is a clear need of maintaining the \textit{honeywords} in $F_P$ \citep{honeyword-juels} \citep{flatness}. But we have found that a large set of defense schemes (identified as M$^\text{FODS}_\text{SOA}$) that provide security against \textit{strong observation attack} (or \textit{strong adversary}) in \textit{fully observable} environment (see detailed in Section \ref{d-cla}) require a direct referral to the original plaintext password for authenticating a legitimate user. Hence to avoid confusion in selecting the original password from the set of \textit{sweetwords} during authentication, these methods cannot afford to store \textit{honeywords} in $F_P$.\\


Mainly motivated by this, we have made the following major contributions in this paper.

\begin{itemize}
\item \textbf{Contribution 1:} We analyze the working principle of M$^\text{FODS}_\text{SOA}$ category methods in detail and show why migration to \textit{honeyword} based scheme is not immediate for a method belonging to this class.


\item \textbf{Contribution 2:} To fill the security gap, we propose few generic principles and show that by following those, \textit{honeywords} can be incorporated to any method belonging to M$^\text{FODS}_\text{SOA}$ class.


\item \textbf{Contribution 3:}  We consider few well known approaches (e.g., S3PAS \citep{strong-ssa-s3pas}, CHC \citep{CHC}, PAS \citep{strong-ssa-pas} and COP \citep{acns}) of M$^\text{FODS}_\text{SOA}$ class and show $-$ by following the proposed principles, how HBAT can be incorporated to those to materialize the proposed idea.
 
\end{itemize}

\textbf{Roadmap:} The rest of the paper is organized as follows. Section \ref{i-g} gives a quick overview of observation attack and classifications of defense strategy. This section also includes a detail analysis to show that with existing setups, why methods belonging to M$^\text{FODS}_\text{SOA}$ class cannot afford to store \textit{honeywords} in $F_P$. Followed by this,  Section\ref{sec:pms} introduces the proposed password masking strategy to guard the original password of M$^\text{FODS}_\text{SOA}$ category methods. Section \ref{sec:ius} then elaborates the security and usability parameters of \textit{honeyword} based approaches which help in determining the overall security and usability standard of a method belonging to M$^\text{FODS}_\text{SOA}$ class after incorporating the HBAT framework. Section \ref{sec:example} shows how the proposed password masking strategy to M$^\text{FODS}_\text{SOA}$ class can be implemented in practice. Finally, the paper is concluded in Section \ref{sec:conclusion} by identifying an open problem in this direction.

\section{Identifying The Gap In $\text{M}^\text{FODS}_\text{SOA}$}
\label{i-g} 
Likewise inversion attack, the threat of shoulder surfing can also significantly affect the security standard of password based schemes. Due to increase in activities of shoulder surfers, in $2002$, the International Standard for PIN Management (ISO 9564) mandates that a PIN entry device should be installed in such a manner so that it can prevent \textit{shoulder surfing attack} (\textbf{SSA}) \citep{bankpin}.  While performing SSA, to obtain the login credentials, an adversary can either record the submitted login information with the help of some recording device (e.g., conceal camera) or, may simply look at the screen/keyboard by standing next to the user. Later, she may use that obtained login credentials to impersonate the genuine user. Depending on the equipments used by an adversary to conduct SSA, the eavesdropper can be classified into following two categories. 
\begin{itemize}
\item \textbf{Strong adversary} uses some recording devices (e.g., conceal camera) to monitor, intercept and analyze each part of a login session \citep{matsumoto-imai-1991} \citep{acns}.

\item \textbf{Weak adversary} relies upon her limited cognitive skills and performs the attack without using any equipment \citep{weak-ssa-bw} \citep{weak-ssa-fc}.
\end{itemize}
 
It is important to note here that if the observation attack is conducted by a strong adversary then it is refereed as strong SSA \cite{strong-ssa-pas}. On the other hand, weak adversary is responsible for performing the weak SSA only \citep{weak-ssa-fc}. Methods belonging to $\text{M}^\text{FODS}_\text{SOA}$ class build security against strong eavesdropper. 


\subsection{Defense strategy and its classifications} 
\label{d-cla}
\textbf{Defense Strategy:} Many methodologies have been developed that can provide security against SSA. In this literature, we identify all these methods as $\text{M}^\text{DS}_\text{OA}$. The $\text{M}^\text{DS}_\text{OA}$ follow a challenge (C) response (R) protocol to avoid the attack. Password (P) can be considered as a shared secret between the user and system. From a set of finite elements, the system first generates a challenge and communicates it to the user. Based on the challenge and password, the user then derives a response and sends it back to the system. Thus, R can roughly be thought of as: R = \textit{f}(C, P). The challenge in each session gets vary and as a consequence, response in each session also gets changed. Therefore without knowing the original password and, just by seeing the passive key entry, it becomes hard for the adversary to derive the actual password.\\

\textbf{Classifications of method:} Based on  how the challenge is being communicated to a user, $\text{M}^\text{DS}_\text{OA}$ can be classified into two categories. The first variation is known as \textit{Partially Observable Defense Scheme} (\textbf{PODS}), where the generated challenge by the system is covertly communicated (may be with the help of an ear-phone) to the user \citep{sssl} \citep{bimodal}. All methods belonging to this category are identified as $\text{M}^\text{PODS}_\text{OA}$. The covert channel in PODS is always assumed to be secure. Thus all the methods belonging to $\text{M}^\text{PODS}_\text{OA}$ category assume that except the user, no one can access to the challenge.

In the latter variation, known as \textit{Fully Observable Defense Scheme} (\textbf{FODS}), the challenge is communicated to the user by using an overt medium \citep{acns} \citep{strong-ssa-pas}. We identify all the methods belonging to this class as $\text{M}^\text{FODS}_\text{OA}$. Therefore, the challenge in $\text{M}^\text{FODS}_\text{OA}$  can be  accessed by all.  

As challenge in PODS is always assumed to be conveyed through a secure channel (between the prover and verifier), thus, in spite of using all the recording gadgets, an adversary fails in retrieving the communicated challenge. This in turn resists the  adversary to derive the original password from the recorded response only. Therefore, all the methods belonging to $\text{M}^\text{PODS}_\text{OA}$ category are capable of defeating the strong adversary. 

In contrast, as challenge is overtly communicated by the methods of $\text{M}^\text{FODS}_\text{OA}$ class, thus an adversary may look at it with/without using a recording device. Hence existing literatures further classify the methods of $\text{M}^\text{FODS}_\text{OA}$ class into two categories $-$ \textit{(a)} $\text{M}^\text{FODS}_\text{OA}$ against strong adversary \citep{strong-ssa-pas}. We identify this as $\text{M}^\text{FODS}_\text{SOA}$ and, \textit{(b)} $\text{M}^\text{FODS}_\text{OA}$ against weak adversary \citep{weak-ssa-fc}. These methods are identified as $\text{M}^\text{FODS}_\text{WOA}$. 

\begin{figure}[!ht]
\centering
\includegraphics[scale=0.7]{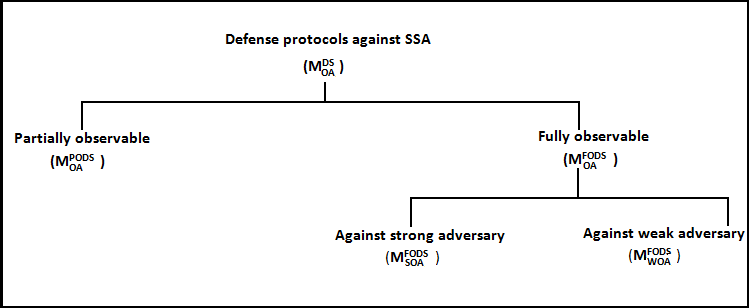}
\caption{Classification of defense mechanisms against SSA.}
\label{classification}
\end{figure}

From the above discussion it is quite intuitive that methods which can defeat strong adversary are also capable of defeating the weak adversary too, but vice versa is not true. In Figure \ref{classification} and Table \ref{note}, we have shown a pictorial overview of the classifications and meaning of the notations related to categorization of methodologies, respectively.

\begin{table}[!h]
\centering
\begin{tabular}{|c|c|}
\hline
Notation & Meaning\\
\hline
$\text{M}^\text{DS}_\text{OA}$ & All the defense schemes against observation attack\\
\hline
$\text{M}^\text{PODS}_\text{OA}$ & Partially observable defense scheme against observation attack\\
\hline
$\text{M}^\text{FODS}_\text{OA}$ & Fully observable defense scheme against observation attack\\
\hline
$\text{M}^\text{FODS}_\text{SOA}$ & Fully observable defense scheme against strong observation attack\\
\hline
$\text{M}^\text{FODS}_\text{WOA}$ & Fully observable defense scheme against weak observation attack\\
\hline
\end{tabular}
\caption{Related notations for categorization of methods}
\label{note}
\end{table}

Remaining of this section unfolds in the following manner:

\begin{itemize}
\item First, we describe the working principle of both $\text{M}^\text{PODS}_\text{OA}$ and $\text{M}^\text{FODS}_\text{WOA}$ and show why methods belonging to these classes are capable of storing \textit{honeywords} with the original password in $F_P$.

\item Followed by this, we elaborate on how the authentication procedure is carried out by the methods belonging to $\text{M}^\text{FODS}_\text{SOA}$ and show why migration to HBAT is not immediate for these methods. 
\end{itemize}



\subsection{Authentication process of $\text{M}^\text{PODS}_\text{OA}$ and $\text{M}^\text{FODS}_\text{WOA}$}
\begin{figure}[!ht]
\centering
\includegraphics[scale=0.6]{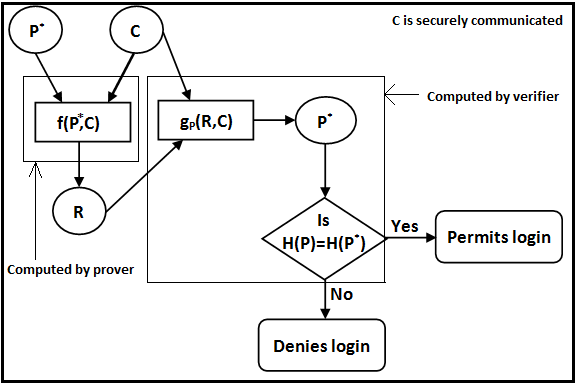}
\caption{Authentication procedure of $\text{M}^\text{PODS}_\text{OA}$.}
\label{PODS}
\end{figure}

In Figure \ref{PODS}, we  have shown the \textbf{authentication procedure of $\text{M}^\text{PODS}_\text{OA}$}. Let P be the shared secret between a user and the system and, this is maintained in $F_P$ by the system. In a session, let the user assumes $\text{P}^*$ as her original secret and performs login based on this assumed token. From the challenge C in that session, the user then generates her response R by using a method f($\text{P}^*, \text{C}$) (ref. to Figure \ref{PODS}). From the submitted R and with the help of C, the verifier then becomes able to generate the same token  P$^*$ (utilized for generating the R) by using a method g$_\text{P}$(C,R). Here g$_\text{P}$() indicates a generator function, used by the verifier to generate the token (here $\text{P}^*$) which has been used by the prover in forming the response. The verifier then checks for a match between the  P$^*$ and the original P. If both P and P$^*$ get matched, the verifier then allows the login otherwise, denies the prover.


Except the fact that the system securely communicates a challenge, \textbf{authentication procedure of $\text{M}^\text{FODS}_\text{WOA}$} follow the same working principle as of $\text{M}^\text{PODS}_\text{OA}$. Methods belonging to this class work on the assumption that an adversary is not allowed to use any computational/recording device to derive the P$^*$ from g$_\text{P}$(C,R). Thus for $\text{M}^\text{FODS}_\text{WOA}$ category methods, the function g$_\text{P}$()  is designed in such a manner so that reverse computation of P$^*$ from g$_\text{P}$(C,R) falls beyond the capacities of limited human cognitive skills \citep{weak-ssa-fc}.\\ 


\textbf{Feasibility of incorporating the honeywords:} From the above discussion, it is understandable that if a user provides her response with reference to a secret then with the help of the communicated challenge, system can uniquely derive that secret from the submitted response. Let both $\text{M}^\text{PODS}_\text{OA}$ and $\text{M}^\text{FODS}_\text{WOA}$ support the HBAT framework where the login information of a user is maintained in the following manner.\\

\begin{center}
\begin{tabular}{c c}
$F_P$: & P$_1$, P$_2$, ..., P$_k$\\
\textit{honeyChecker}: & t
\end{tabular}
\end{center} 

where P$_i$ denotes a password maintained at $i^{th}$ index and, P$_t$ and \textit{t} ($1 \leq t \leq k$) correspond to the original password and index position of the original password, respectively. 

From the compromised $F_P$, if the eavesdropper gets this list (P$_1$, P$_2$, ..., P$_k$) and guesses P$_i$ ($1 \leq i \leq k$) as the original password, then from the submitted response by the attacker, the system will uniquely be able to derive that P$_i$ by using the generator function g$_\text{P}$(). This in turn makes system understand that the login has been performed by using the secret stored at $i^{th}$ index. System then communicates this index value to the \textit{honeyChecker} and follows the normal HBAT authentication routine to detect the threat. This infers that both $\text{M}^\text{PODS}_\text{OA}$ and $\text{M}^\text{FODS}_\text{WOA}$ can readily support HBAT to reinforce their security standard.


\subsection{Authentication process of $\text{M}^\text{FODS}_\text{SOA}$}
\label{ap-fods}
\begin{figure}[!ht]
\centering
\includegraphics[scale=0.6]{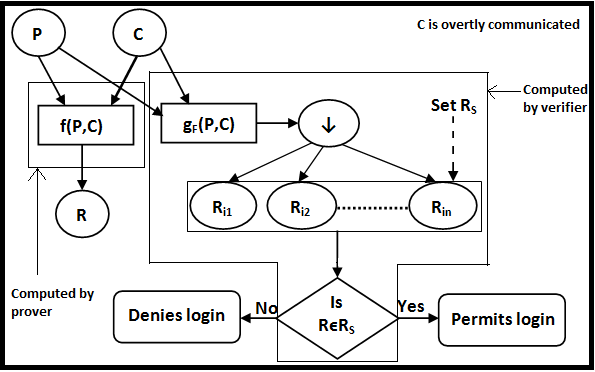}
\caption{Authentication procedure in $\text{M}^\text{FODS}_\text{SOA}$; ``$\downarrow$" inside the circle denotes mapping operation. $\text{R}_\text{ik}$ implies $k^{th} (\geq 1)$ instance of the valid response in the set $R_S$ generated by g$_\text{F}$(P,C) in the $i^{th}$ round.}
\label{FODS}
\end{figure}

Methods fall under this category are designed to defeat the strong adversary where the challenge is overtly communicated over an insecure channel. The adversary here may use recording gadgets to monitor, intercept and analyze each part of an authentication session. Therefore, if authentication procedure in $\text{M}^\text{FODS}_\text{SOA}$ follow the same path as of $\text{M}^\text{PODS}_\text{OA}$ (or $\text{M}^\text{FODS}_\text{WOA}$) then from the recorded challenge C (communicated by the system) and the response R (submitted by user), the adversary can easily derive the original password uniquely by using the same function g$_\text{P}$(C,R). Hence, $\text{M}^\text{FODS}_\text{SOA}$ follow a different strategy to defeat the powerful adversary.  In Figure \ref{FODS}, we have shown the authentication procedure followed by $\text{M}^\text{FODS}_\text{SOA}$ category methods.

Like PODS, after receiving the challenge, the user generates a response in the similar fashion. To validate the response provided by the user, a different method namely, g$_\text{F}$(P,C) is used by $\text{M}^\text{FODS}_\text{SOA}$. Below we gist characteristic of the generator function g$_F$()

\begin{itemize}
\item As inputs, it takes the original secret of user (P) maintained in $F_P$ and the challenge (C) communicated by the system.

\item From those inputs, it generates a set of probable response elements. The set may contain a single (e.g., COP described in Section \ref{cop}) or multiple elements (e.g., S3PAS described in Section \ref{s3pas}). We identify this as response set ($R_S$).
\end{itemize}  

If the submitted response by user belongs to $R_S$ then only $\text{M}^\text{FODS}_\text{SOA}$ allows the login. From the response submitted by the user and the communicated challenge by the system, an adversary always derives multiple probable secrets and hence cannot derive the original secret uniquely before recording a certain number ($> 1$) of authentication sessions (or challenge-response pairs). \\

 To defeat a strong adversary for more number of authentication sessions, most of the existing methods of $\text{M}^\text{FODS}_\text{SOA}$ class follow complex login strategies . Therefore to explain the working principle of $\text{M}^\text{FODS}_\text{SOA}$ with an example, we have designed a hypothetical approach which holds all properties of $\text{M}^\text{FODS}_\text{SOA}$ category methods though follows a simple authentication procedure.\\

\begin{figure}[!ht]
\centering
\includegraphics[scale=0.7]{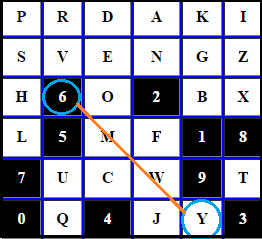}
\caption{Visual interface of the hypothetical method belonging to $\text{M}^\text{FODS}_\text{SOA}$ class. The virtual line connecting the characters $6$ and \textit{Y} is shown in the above figure.
\label{hypo}}
\end{figure}

\textbf{Hypothetical FODS (an example):} Let a set of $36$ elements (digits and alphabets) be used to design the hypothetical method. The user selects an alphanumeric password of length $4$ for authentication. In a session, all the $36$ elements from the alphanumeric set are randomly arranged in a $6 \times 6$ matrix and are displayed on the login screen. This matrix acts as a challenge in this hypothetical system and in Figure \ref{hypo}, we have shown one instance of this.

We assume that an authentication session in this hypothetical system be of $4$ login rounds. In order to give response to a challenge, in each login round, the user first selects two consecutive characters of her password (in a cyclic manner) and locates those characters on the matrix. Then user connects those two characters by means of a virtual line and selects a character belonging to that line as response.

Let the password chosen by the user be \textit{6YJC}. Therefore in order to generate the responses, the user selects \textit{6Y} in the first round, \textit{YJ} in the second round, \textit{JC} in the third round and \textit{C6} in the last round. System also selects the password by parts to generate the $R_S$ with reference to the challenge matrix shown in Figure \ref{hypo}. For example, in the first round, as both \textit{M} and \textit{W} belong to the straight line connecting \textit{6} and \textit{Y}, thus, system forms the $R_S$ as $\{$\textit{M, W}$\}$ in the first round. In other words, g$_\text{F}$() returns the response set as $\{$\textit{M, W}$\}$ in the first round. To pass the first authentication round, the user may enter either of these elements as response. 
The login procedure for remaining rounds follow the same path with changes in the elements of $R_S$ and the user response.

\subsection{Identifying the gap}
\label{gap}
The discussion from the previous section infers that methods belonging to $\text{M}^\text{FODS}_\text{SOA}$ class require a reference to the original password to build the response set $R_S$ which in turn validates the submitted response by a user. Let a method of $\text{M}^\text{FODS}_\text{SOA}$ class supports HBAT framework and, maintains $k-1$ ($k > 1$) \textit{honeywords} along with the original password in $F_P$. 

Now, to conduct an authentication session, system first generates a challenge. Thereafter, in order to generate the $R_S$, the function  g$_\text{F}$() requires both communicated challenge and the original password. Though g$_\text{F}$() can readily avail the challenge, but gets confused among \textit{k} \textit{sweetwords} while selecting the original password for formation of $R_S$. As output of g$_\text{F}$() changes with change in any of its parameters, thus selecting any \textit{honeyword} instead of the original password will produce a different response set which may deny the correct response of a legitimate user. To avoid such confusion $-$ in distinguishing the original password from the set of \textit{honeywords}, any method belonging to $\text{M}^\text{FODS}_\text{SOA}$ class cannot afford to support HBAT. 

It is \textbf{very important} to note here that though each \textit{sweetword} will generate a different $R_S$ with respect to a fixed challenge, but there may exist some overlapping among the elements of $R_S$. For example, for  two different \textit{sweetwords} \textit{6Y} and \textit{LF}, the generated $R_S$ with respect to the challenge shown in Figure \ref{hypo} will be $\{$M, W$\}$ and $\{$5, M$\}$, respectively. This shows that though generated $R_S$ are different for two different \textit{sweetwords}, but they share a common element \textit{M}.

\subsection{Out of scope threat}
We do not attempt to address password compromise due to SSA by an adversary who already obtains the password file. Even if the original password is guarded by the \textit{honeywords}, by seeing the response, the adversary will always identify the original password. Building a defense mechanism which can address such attack is an open problem till date.


\section{Proposed Password Masking Strategy}
\label{sec:pms}
As we discuss the authentication procedure of $\text{M}^\text{FODS}_\text{SOA}$ category methods under the light of a hypothetical scheme in Section \ref{ap-fods}, most of the existing schemes of $\text{M}^\text{FODS}_\text{SOA}$ class display the challenge on the login terminal \citep{yan-2015}. In order to generate the response, user then finds appropriate clue from the challenge with reference to her original password. To verify the submitted response, system first generates \textit{a set of valid responses} ($R_S$) with reference to the challenge and the original password of the user.  If the provided response by the user belongs to $R_S$ then system validates the user otherwise, denies the login.\\

While identifying the existing security gap of the methods belonging to $\text{M}^\text{FODS}_\text{SOA}$ class in Section \ref{gap}, we have made a very important point there. Though there may exist some overlapping, but with reference to a challenge, each \textit{sweetword} creates a different $R_S$. The core of this contribution is mainly influenced by this. As function g$_\text{F}$() in $\text{M}^\text{FODS}_\text{SOA}$ takes two parameters (password and challenge) to generate the $R_S$, therefore in order to fit $k-1$ \textit{honeywords} in $\text{M}^\text{FODS}_\text{SOA}$, we will adopt the following strategy.

\begin{itemize}
\item System first selects $k$ different \textit{sweetwords} in such a manner so that given any of some particular challenges, they are capable of generating $k$ mutually exclusive (non-overlapping) $R_S$ by using the generator function g$_\text{F}$().

\item With respect to those \textit{sweetwords}, system then generates a challenge in such a way so that produced $R_S$ by each of $k$ \textit{sweetwords} are mutually exclusive.

\item As response sets are mutually exclusive, thus, if provided response element by a user belongs to a particular $R_S$ then that will not be a member of any other $R_S$.

\item Therefore, from the response submitted by the user, system can uniquely identify the corresponding $R_S$, which in turn makes system understand the corresponding \textit{sweetword} responsible for creating that $R_S$.   

\item Finally, with reference to the index of that \textit{sweetword}, by communicating with the \textit{honeyChecker}, system can detect the breach.

\end{itemize}

As above method looks promising for incorporating the \textit{honeywords} to the methods belonging to $\text{M}^\text{FODS}_\text{SOA}$ class, thus we have identified few factors (e.g., selection of \textit{honeywords}, principle behind generating the challenge) that will play a crucial role to materialize this concept.

\subsection{Principles for generating the challenge and honeywords} 
\label{p-c-h}
Let the algorithm used by the $\text{M}^\text{FODS}_\text{SOA}$ (without incorporating \textit{honeywords}), for carrying out the authentication procedure be denoted by  $\mathcal{A}$. With the existing setup, therefore $\mathcal{A}$ only generates a single $R_S$ with reference to a visual challenge and the password maintained in $F_P$. But to adopt the HBAT framework, a visual challenge should be generated in such a manner so that \textit{k} \textit{sweetwords}, with reference to the visual challenge, can form \textit{k} mutually exclusive (or non-overlapping) $R_S$. Hence to support HBAT, some modifications are required to  $\mathcal{A}$. We identify the modified algorithm  as $\mathcal{A}^+$.
Formation of \textit{k} non-overlapping $R_S$ is mainly driven by two components $-$ the visual challenge and  \textit{sweetwords}. In this section, we mainly focus on how these two collaborate to incorporate HBAT framework in $\text{M}^\text{FODS}_\text{SOA}$. \\

During the discussion of the working principle of $\text{M}^\text{FODS}_\text{SOA}$ category methods under the light of a hypothetical system in Section \ref{ap-fods}, we show that system (and user) may consider a substring of the password to conduct each round of an authentication session. Let for a password of length $\ell (> 0)$, a method of $\text{M}^\text{FODS}_\text{SOA}$ class takes \textit{lr} ($> 0$) login rounds to complete an authentication session. Next, we define two important factors that play very crucial roles in incorporating HBAT to $\text{M}^\text{FODS}_\text{SOA}$ category methods.\\

\textbf{Definition 1:} \textbf{Partial Password Information:} \textit{Partial password information or PPI is the smallest sub part of the password accepted by} g$_\text{F}$() \textit{to conduct each round of an authentication session.}\\

For a password of length $\ell$, the length of \textit{PPI} may vary between $1$ and $\ell$. For example, g$_\text{F}$() in S3PAS  \citep{strong-ssa-s3pas} always accepts a \textit{PPI} of length lesser than $\ell$ whereas, g$_\text{F}$() in COP \citep{acns} accepts the whole password as \textit{PPI}. \\

\textbf{Definition 2:} \textbf{Partial Response Set:} \textit{The response set returned by the} g$_\text{F}$(), \textit{which accepts a PPI and the challenge as parameters, is identified as partial response set or PRS.}\\

For example, for the \textit{PPI} as \textit{6Y}, the g$_\text{F}$() of the hypothetical system in Section \ref{ap-fods}, will return the \textit{PRS} as $\{$M, W$\}$  with reference to the visual challenge shown in Figure \ref{hypo}.\\

\textbf{Note 1:} The basic motivation behind modifying the $\mathcal{A}$ to  $\mathcal{A}^+$ is to incorporate HBAT framework to $\text{M}^\text{FODS}_\text{SOA}$. With $\mathcal{A}^+$ running at back end, let us assume that a method belonging to  $\text{M}^\text{FODS}_\text{SOA}$ class is maintaining $k$ \textit{sweetwords}. In the $i^{th}$ round, let \textit{PPI} and \textit{PRS} corresponding to the $j^{th}$ ($1 \leq j \leq k$) \textit{sweetword} be denoted as \textit{PPI}$^{j}_{i}$ and \textit{PRS}$^{j}_{i}$, respectively. With reference to the challenge C, g$_\text{F}$(\textit{PPI}$^{j}_{i}$, C) then returns  
\textit{PRS}$^{j}_{i}$ in the $i^{th}$ round. 
Now, for \textbf{\textit{k} mutually exclusive \textit{PRS} in each round}, if submitted response by a user always belongs to a \textit{PRS}, formed by the \textit{PPI} of $t^{th}$ \textit{sweetword}, then a method of $\text{M}^\text{FODS}_\text{SOA}$ class understands that the user has performed login against the $t^{th}$ \textit{sweetword}.\\

\textbf{Note 2:} Generating a visual challenge which will form \textit{k} mutually exclusive \textit{PRS} in each round may not be a feasible solution as this may consume a huge amount of time. Therefore, a more feasible solution will be to generate a visual challenge in such a way that can generate \textbf{\textit{k} mutually exclusive \textit{PRS} atleast in a single round}. Based on this, we first propose the following principle.\\

\textbf{\textit{Principle 1:}} \textit{During authentication, a method belonging to $\text{M}^\text{FODS}_\text{SOA}$ class should maintain \textit{k} ($\geq 2$) \textit{sweetwords} in such a manner so that atleast in a round (say $i^{th}$ round: where $1 \leq i \leq lr$), $\mathcal{A}^+$ will generate a visual interface (challenge) on which  \textit{PRS}$^{1}_{i}$, \textit{PRS}$^{2}_{i}$ $...$  \textit{PRS}$^{k}_{i}$ will be mutually exclusive}.\\

\textbf{Discussion 1:} This principle infers how challenge should be generated by $\text{M}^\text{FODS}_\text{SOA}$ category methods. According to the above principle, except in a particular round, say $i$, generated $k$ \textit{PRS} by the \textit{PPI} of \textit{sweetwords} may get overlapped.  Therefore, except in $i^{th}$ round, system may find that provided response by a user is belonging to multiple \textit{PRS}. But in the $i^{th}$ round, if the submitted response belongs to $t^{th}$ $(1 \leq t \leq k)$ \textit{PRS} (formed by the $t^{th}$ \textit{PPI}) then that must not belong to any other \textit{PRS}. Not only in $i^{th}$ round, but response in each round if maps to a \textit{PRS} formed by the $t^{th}$ \textit{PPI}, then only the system considers that the login has been performed against $t^{th}$ \textit{sweetword}. The system then directs the index value \textit{t} (along with username) to the \textit{honeyChecker} server and performs normal HBAT routine to detect the breach.\\

Other than the visual challenge, the selection strategy of \textit{sweetwords} are equally important for incorporating the HBAT framework to $\text{M}^\text{FODS}_\text{SOA}$ as g$_\text{F}$() takes both the challenge and \textit{sweetword} as inputs to carry out an authentication session. Therefore, next principle focuses on how the \textit{sweetwords} are needed to be chosen  by a method of $\text{M}^\text{FODS}_\text{SOA}$ class.\\

\textit{\textbf{Principle 2:}} \textit{Not only the sweetwords, but in each round,  PPI of the \textit{sweetwords} must also be different}.\\

\textbf{Discussion 2:} Let  \textit{6YJC} and \textit{6YM5} be the two \textit{sweetwords} maintained by the hypothetical system belonging to  $\text{M}^\text{FODS}_\text{SOA}$ (ref. to Section \ref{ap-fods}). Though these two \textit{sweetwords} are different, but \textit{PPI} of those in the first round are same (i.e., \textit{6Y}). Now if the $\mathcal{A}^+$ randomly chooses first round to generated $2$ non-overlapping \textit{PRS} then it would not be possible as both the \textit{PPI} are same. Thus for an algorithm, dynamically selects a round to generate \textit{k} mutually exclusive \textit{PRS}, must follow the above principle while selecting the \textit{sweetwords}.\\

During incorporation of HBAT framework to $\text{M}^\text{FODS}_\text{SOA}$, the value of \textit{k} plays a major role to protect the original password. As \textit{k} increases, it creates more confusion in a attacker's mind. Thus we state the following principle which guides any method of $\text{M}^\text{FODS}_\text{SOA}$ class for choosing a suitable value of \textit{k}.\\

\textit{\textbf{Principle 3:}} \textit{Number of sweetwords (i.e., value of k) varies depending on a method belonging to $\text{M}^\text{FODS}_\text{SOA}$ class and it varies between $2$ and number of all possible responses}.\\

\textbf{Discussion 3:} Let the number of all possible responses that can be generated by a user be $\mathcal{Z} (\geq 2)$. Let the number of average response elements belonging to each \textit{PRS} be $\mathcal{E} (\geq 1)$. If \textit{k} \textit{PRS} in a round are capable of covering all the response elements, then the following equation stands

\begin{equation}
k = \dfrac{\mathcal{Z}}{\mathcal{E}}
\end{equation} 

Now as value of $\mathcal{Z}$ varies depending on a method of  $\text{M}^\text{FODS}_\text{SOA}$ category (e.g., $\mathcal{Z} = 94$ for S3PAS \citep{strong-ssa-s3pas} and that of $\mathcal{Z} = 4$ for PAS \citep{strong-ssa-pas}) so does $\mathcal{E}$. Thus the ratio of these two components, \textit{k}, also varies.\\

It is important to note here that to create illusion in attacker's mind a system must store atleast one \textit{honeyword} along with the original password. Thus  \textit{k} always gets a minimum value $2$. On the other hand, as there will be atleast one element in each \textit{PRS}, hence maximum value of \textit{k} can be reached upto $\mathcal{Z}$.\\

\subsection{Proof of the concept}
In this section, we will show that aforementioned $3$ principles are sufficient for incorporating HBAT to any method belonging to $\text{M}^\text{FODS}_\text{SOA}$ class.\\

 Obeying their \textbf{existing authentication routine}, following things stand for any method of $\text{M}^\text{FODS}_\text{SOA}$ class.

\begin{center}
\begin{tabular}{c c c}
g$_\text{F}$(P,C) $\longrightarrow$ $R_S$ & and  & f(P,C) $\longrightarrow$ R

\end{tabular}
\end{center}

and,
\[
    \text{In verification phase} \Longrightarrow
\begin{cases}
     \text{Allows prover } & \text{if } R \in R_S \\        \text{Denies the login},         & \text{otherwise}
\end{cases}
\]

By following the \textbf{modified framework}, let a method of $\text{M}^\text{FODS}_\text{SOA}$ category chooses $i^{th}$ ($1 \leq i \leq lr$) round for distinguishing original password from the set of $k$ \textit{sweetwords}. As already shown previously, the value of $k$ can be reached maximum upto $\mathcal{Z}$ and cannot be no lesser than $2$ (see the proof of \textit{Principle} $3$ in the previous section). We denote $j^{th}$ ($1 \leq j \leq k$) \textit{sweetword} as $\mathcal{S}_j$ and \textit{PPI} of it in $i^{th}$ round is identified as $\text{PPI}_{i}$($\mathcal{S}_j$). Let the response element corresponding to $j^{th}$ \textit{sweetword} in the $i^{th}$ round be denoted by $r^j_i$.
Therefore in the $i^{th}$ round, following equation must stand to eliminate the ambiguity

\begin{equation}
r^1_i \neq ...\neq r^{k-1}_i \neq r^k_i
\label{eq-pf-1}
\end{equation} 

For those methods, generating $R_S$ of single element, the following equation then stands immediately

\begin{equation}
^{i}R_S^1 \neq ... ^{i}R_S^{k-1} \neq ^{i}R_S^{k}
\label{eq-pf-2}
\end{equation}  
where $^{i}R_S^j$ denotes a response set generated by $j^{th}$ \textit{sweetword} in the $i^{th}$ round.\\

As proposed principles here are intended to cover all the methods belonging to $\text{M}^\text{FODS}_\text{SOA}$ class, therefore, authentication protocols that generate $R_S$ having more than one element must also need to satisfy the above equation.\\

\textbf{Proposition 1:} Above discussion implies that in a round, there is a clear need of generating $k$ mutually exclusive response sets for incorporating HBAT to any method of $\text{M}^\text{FODS}_\text{SOA}$ class.\\

Equation \ref{eq-pf-2} also complements the fact which can be described in the form of the following equation

\begin{equation}
\text{g}_\text{F}(\text{PPI}_{i}(\mathcal{S}_1),\text{C}) ... \neq \text{g}_\text{F}(\text{PPI}_{i}(\mathcal{S}_{k-1}), \text{C}) \neq \text{g}_\text{F}(\text{PPI}_{i}(\mathcal{S}_k), \text{C}) 
\label{eq-pf-3}
\end{equation}

\textbf{Proposition 2:} As Equation \ref{eq-pf-3} shows that response set returned by each $\text{g}_\text{F}(\text{PPI}_{i}(\mathcal{S}_j),\text{C})$ are unique, therefore for the fixed challenge value C, PPI of each \textit{sweetword} must be different.\\

\textbf{Proposition 3:} Also, as g$_\text{F}$() takes C as one of the parameters, therefore, both Equation \ref{eq-pf-2} and Equation \ref{eq-pf-3} suggest that generated C by the system plays a huge role in generating $k$ non-overlapping $R_S$.\\

\textbf{Note 3:} As system selects the value of $i$ randomly, therefore proposition 2 proofs the \textit{Principle} $2$. Also, proposition $1$ along with proposition $3$ validates \textit{Principle} $1$. In Section \ref{p-c-h}, we already proof the third principle and hence, we may claim that proposed principles here are sufficient for incorporating HBAT to any method belonging to  $\text{M}^\text{FODS}_\text{SOA}$ class. \\

In Section \ref{sec:example} too, we show that by following \textit{Principle} $1$ to \textit{Principle} $3$, any method belonging to $\text{M}^\text{FODS}_\text{SOA}$ class can afford to store \textit{honeywords} in practice. As the existing underlying algorithm ($\mathcal{A}$) needs to be modified (to $\mathcal{A}^+$) for incorporating the HBAT, thus it may impact the whole authentication procedure. Next we discuss the impact of this modification on $\text{M}^\text{FODS}_\text{SOA}$.

\subsection{Impact of modification from $\mathcal{A}$ to $\mathcal{A}^{+}$}
The impact of the modification may cause some changes either in system's behaviour or, in user's behaviour or, both. By adopting the $\mathcal{A}^{+}$, a system may change its visual (login) interface and/or, a user may require to remember some additional information for login. From now onwards, we denote \textit{change in system's login interface} as \textbf{CSLI} and \textit{remembering an extra  information} by the user as \textbf{REI}. As HBAT can be considered as an extended version of the password based authentication, thus a balance is required between the security and usability factors. We have analyzed the effect of the modification from both security and usability perspectives.

\subsubsection{Impact of the modification from security perspective} 
\label{msp}   
During modification from $\mathcal{A}$ to $\mathcal{A}^{+}$, basic security standard (includes security against observation attack through brute force search, password guessing attack \citep{password-guess-attack} etc.)  of $\mathcal{A}$ can be influenced in one of the following three ways $-$
\begin{itemize}
\item[1.]  $\mathcal{A}^{+}$ may provide enhanced security compared to $\mathcal{A}$.
\item[2.] $\mathcal{A}^+$ may provide same security as of $\mathcal{A}$.
\item[3.] $\mathcal{A}^{+}$ may provide lesser security than $\mathcal{A}$.
\end{itemize}
During modification to $\mathcal{A}^+$, the target should always be to ensure higher security standard (if possible) or atleast the same security standard as of $\mathcal{A}$.

\subsubsection{Impact of the modification from usability perspective}
\label{mup}
\textit{CSLI} may influence the login procedure. We identify the \textit{change in login procedure} as \textbf{CLP}. Modification to $\mathcal{A}^{+}$ may also cause \textit{REI} by a user. Based on \textit{CSLI}, a prover may have to face any one of the following two situations during login $-$ (a) \textit{CSLI} with \textit{CLP} and (b) No \textit{CSLI} with no \textit{CLP}. It is also noticeable that \textit{CLP} will cause \textit{REI} by a prover and vice versa is also true. On the other hand, no \textit{CLP} will cause no \textit{REI} by  a user. Now by combining these three components $-$ \textit{CSLI}, \textit{REI} and \textit{CLP}, the overall usability standard of $\mathcal{A}^{+}$ can fall under one of the following  categories $-$

\begin{itemize}
\item[1.] \textit{CSLI} with \textit{CLP} and \textit{REI}.
\item[2.] No \textit{CSLI} with no \textit{CLP} no and \textit{REI}. 
\end{itemize}

While modifying to $\mathcal{A}^{+}$, desired objective will be achieving no \textit{CLP} and no \textit{REI} from the user end.

\section{Integrated Usability and Security Features}
\label{sec:ius}
HBAT framework has its own security and usability parameters. While incorporating HBAT to $\text{M}^\text{FODS}_\text{SOA}$, then it inherits the usability and security parameters of HBAT.

\subsection{HBAT usability features}
\label{huf}
There are three well defined usability features \citep{honeyword-juels}  related to a HBAT $-$ \textit{(a) System interference}, \textit{(b) Stress on memorability} and \textit{(c) Typo safety}.\\

\textbf{(a) System interference:} During registration, some HBAT (like \textit{take-a-tail} \citep{honeyword-juels}) require an extra information to be remembered by a user. This extra information is used as part of user's login credential. In this situation, as a HBAT interferes on the password choice of user, thus this property is known as \textit{system interference}. If a user needs to remember \textit{n} system generated information for \textit{n} different login accounts then \textit{system interference} becomes high (e.g., \textit{take-a-tail} \citep{honeyword-juels}). If a HBAT allows a user to choose the extra information then user can use that information for login into \textit{n} different accounts. This is identified as \textit{low system interference} (e.g., \textit{modified-tail} \citep{few-honeywords}). There are some HBAT too which do not force users to remember any extra information \citep{kamouflage}.  A HBAT, having \textit{high system interference}, does not provide good usability standard. \\

\textbf{(b) Stress on memorability:} There exists a relationship between the system interference and stress on memorability. If a \textit{honeyword} based scheme imposes high system interference, then a user has to remember different system generated information for different login accounts. This increases the stress on user's mind and threats the usability standard. On the other hand, user may use the same login credential for different login accounts using a HBAT having low or, no system interference. This causes low stress on memorability which is a desirable criteria for a HBAT to be user friendly.\\

\textbf{(c) Typo safety:} A \textit{honeyword} generation algorithm is called typo safe if typing mistake of a user rarely matches with any \textit{honeyword}, maintained against that user account. A less typo safe method, due to typing mistake of the legitimate user, may mislead the \textit{hoenyChecker} in generating a negative feedback signal even though $F_P$ has not been compromised.

\subsection{HBAT security features}
\label{hsf}
There are three well defined security parameters related to a HBAT $-$ \textit{(a) Denial of Service (DoS) resistivity}, \textit{(b) Defending Multiple system vulnerability (MSV)} and \textit{(c) Flatness}. \\

\textbf{(a) DoS resistivity:}  Without compromising $F_P$, if an adversary becomes able to guess any \textit{honeyword} then the \textit{honeyword} based approach is identified as weak DoS resilient one \citep{few-honeywords}. There are few \textit{honeyword} generation techniques which help adversary to guess a \textit{honeyword} easily if the original password is known to her. In such situation, an attacker may intentionally submit a \textit{honeyword} to make system understand that $F_P$ has been compromised. Once system senses submission of a \textit{honeyword}, it blocks each user account and denies any further login attempt. To mount DoS attack, an attacker needs to know the original password of a user. To get the original password, adversary may either perform observation attack or, may create her own account.

Though above discussion is fruitful to give a generic idea about how DoS attack is performed, but the nature of the attack slightly gets changed while incorporating HBAT to $\text{M}^\text{FODS}_\text{SOA}$. In a particular round (say $i: 1 \leq i \leq lr$), to identify a \textit{sweetword} uniquely, $\mathcal{A}^+$ maps each \textit{sweetword} to a unique $R_S$. Therefore, if an adversary can successfully guess both the $i$ and an element of a $R_S$, formed by the \textit{PPI} of a \textit{honeyword}, then she can intentionally submit that element to facilitate her chances in mounting the DoS attack. Particularly, for a small value of $lr$ and $k \approx \mathcal{Z}$, chances of mounting this attack become higher. 

Therefore, if possible, with providing moderate threat detection rate, the value of $k$ should be chosen in such a manner so that the response sets generated by the \textit{sweetwords} do not cover all possible response elements. For the systems provide weak security against DoS, light security policy against DoS \citep{honeyword-erguler} may be adopted to defend the attack. Under light security policy, a system only blocks that account against which it senses submission of a \textit{honeyword}.\\

\textbf{(b) Defending  MSV:} Recent reports strongly suggest that normal users mostly select same passwords for multiple accounts \citep{tangled} \citep{password-choosing-tendency}. During registration, after receiving password from a user, system generates a list of \textit{honeywords}. One of the characteristics of any \textit{honeyword} generation approach is to generate different list of \textit{honeywords} at different run of the algorithm even though the password remains same. Eventually, the lists of \textit{sweetwords} in two different systems differ even if the same password is used. 

Now if an adversary becomes successful in obtaining $F_P$ from both these systems then there is a high probability that intersection between two lists of \textit{sweetwords} will reveal the original password of the user. This is known as Multiple System Vulnerability or, MSV crisis of HBAT.\\

\textbf{(c) Flatness:} From the compromised $F_P$, maintaining \textit{honeywords} along with the original password, an attacker gets confused among \textit{k} \textit{sweetwords} for each user account. Now sometimes it may happen that the adversary can easily identify the original password of a user from the list of \textit{sweetwords} (e.g., if there exists a correlation between the username and password). A HBAT is said to be \textit{perfectly flat} if an adversary has no advantage in identifying the original password of the user from the list of \textit{sweetwords}. Thus, for a \textit{perfectly flat} \textit{honeyword} generation algorithm, the probability of selecting the original password  becomes $1/k$. 
For an \textit{approximately-flat} HBAT, the probability of  selecting the original password from the list of \textit{sweetwords} becomes slightly higher than $1/k$. A good \textit{honeyword} generation algorithm is required to be \textit{perfectly-flat} to obfuscate the attacker properly.\\

While incorporating \textit{honeywords} to the $F_P$ of a method belonging to $\text{M}^\text{FODS}_\text{SOA}$ class, it is important to analyze the security and usability aspects associated with the HBAT framework. Next we show how the concept of password masking strategy can be incorporated to some of the methods belonging to $\text{M}^\text{FODS}_\text{SOA}$ category in practice.

\section{Incorporating Honeywords to $\text{M}^\text{FODS}_\text{SOA}$}
\label{sec:example}

To the best of our belief, long back in $1991$, proposed concept of Matsumoto and Imai \citep{matsumoto-imai-1991} can be considered as the very first contribution in the domain of $\text{M}^\text{FODS}_\text{SOA}$ category methods. Researchers have introduced many methodologies (belonging to $\text{M}^\text{FODS}_\text{SOA}$ class) thereafter \citep{fods-01} \citep{CHC} \citep{ssa-2006} \citep{acns}. Among them, while some provide really good security against powerful eavesdropper with lower usability standard \citep{strong-ssa-hopper} \citep{ssa-2006}, the others, come with good usability standard though lagging in providing good security against strong adversary. In this section, we focus on some of the methods that maintain a nice balance between the security and usability aspects. Therefore, to show that proposed idea is implementable in practice, the methods  we have considered here are --- S3PAS \citep{strong-ssa-s3pas}, CHC \citep{CHC}, PAS \citep{strong-ssa-pas} and COP \citep{acns}. Next, we consider these methods one by one and incorporate HBAT to those by following the proposed principles in Section \ref{p-c-h}.

\subsection{Incorporating Honeywords to S3PAS}
\label{s3pas}
In this section, we first give a quick overview of the existing working principle of S3PAS \citep{strong-ssa-s3pas} belonging to $\text{M}^\text{FODS}_\text{SOA}$.  Then we show how HBAT framework can be integrated with this existing approach.\\

\subsubsection{Working principle of S3PAS}  
In basic S3PAS, from a character set of \textit{T} elements, a user first selects a password of length $4$. In each session, \textit{T} characters are randomly arranged in a $m \times n$ matrix and displayed on a visual interface. The orientation of characters in the $m \times n$ matrix remains static for an entire session.

\begin{figure}[!h]
\centering
\includegraphics[scale=0.34]{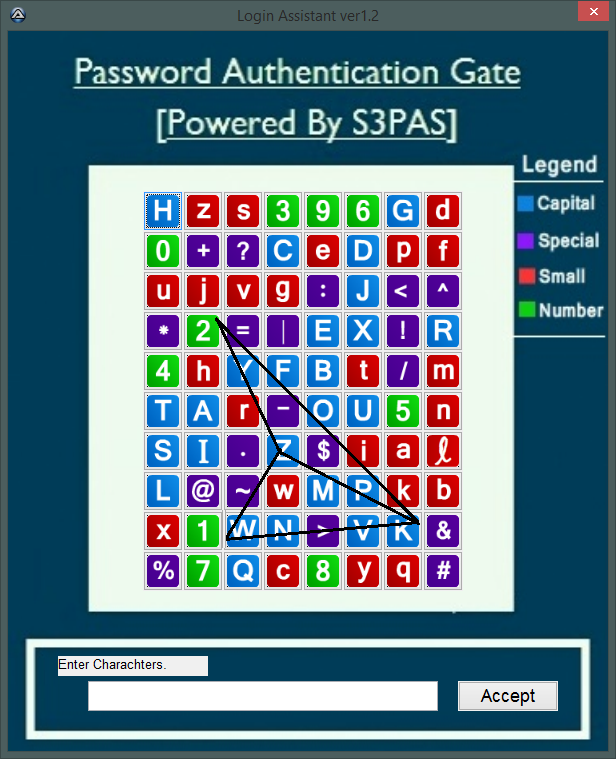}
\caption{Visual interface of S3PAS with $T = 80$. Responses in first two rounds for the password \textit{2KZW}. The formed virtual triangles (marked by black border) in the first two rounds are $\bigtriangleup$ \textit{2KZ} and $\bigtriangleup$ \textit{KZW}. The valid responses may be $``-"$ for $\bigtriangleup$ \textit{2KZ} and \textit{$``$M$"$} for $\bigtriangleup$ \textit{KZW}. 
\label{resp-s3pas}}
\end{figure}    

S3PAS is comprising of $4$ login rounds in each session. Starting from $i^{th}$ index of the password, in each round \textit{i} (from $1$ to $4$), the prover selects $3$ consecutive characters of her password in a cyclic manner. 
On the visual interface, the prover then creates a virtual triangle by using three consecutive password characters correspond to that round. To submit her response, the user selects any character that falls under the area of the formed virtual triangle in that round. 
Let the password of a user be \textit{2KZW}. Therefore formed virtual triangles are $\bigtriangleup$ \textit{2KZ} in $1^{st}$ round, $\bigtriangleup$ \textit{KZW} in $2^{nd}$ round, $\bigtriangleup$ \textit{ZW2} in $3^{rd}$ round and $\bigtriangleup$ \textit{W2K} in $4^{th}$ round. In Figure \ref{resp-s3pas}, we have shown the visual interface of S3PAS and formed virtual triangles in first two rounds.\\

\subsubsection{Proposed modification for incorporating HBAT framework} By following principle $1$ (in Section \ref{p-c-h}), atleast in a round, there must be no common elements among the \textit{PRS} of \textit{sweetwords}. In S3PAS, as elements inside a triangle form a \textit{PRS}, thus, \textit{honeywords} need to be selected (and challenge should be created) in such a manner so that there exist no overlapping areas among the formed triangles in a particular round. To ensure this, verifier selects \textit{k} \textit{sweetwords} in such a manner so that formed triangles (or \textit{PPI} of \textit{sweetwords}) in any round are different (ref. to principle $2$).

 Therefore, for the password \textit{2KZW}, generated \textit{honeywords} (for \textit{k} as $6$) may be \textit{8IMN, 6ABS, 0XRJ, 3OVB} and \textit{rD1$\ell$}; as for any login round, \textit{PPI} of these \textit{sweetwords} are different. For example, without loss of generality, the \textit{PPI} of \textit{sweetwords} in the $3^{\text{rd}}$ round are \textit{ZW2},   \textit{MN8},  \textit{BS6},   \textit{RJ0},   \textit{VB3}. and   \textit{1$\ell$r}.\\

Let the background algorithm of existing S3PAS scheme be denoted by $\mathcal{A}_\text{S3PAS}$. We denote the modified algorithm that supports HBAT feature as $\mathcal{A}^+_\text{S3PAS}$. The working principle of $\mathcal{A}^+_\text{S3PAS}$ is following.

\begin{itemize}
\item \textbf{Step 1:} $\mathcal{A}^+_\text{S3PAS}$ first randomly selects a round for generating $k$ non-overlapping triangles    by the \textit{PPI} of \textit{sweetwords} on the visual interface. Let the selected round be $i$: where ($1 \leq i \leq 4$).

\item \textbf{Step 2:} Verifier randomly allocates \textit{T} characters to $m \times n$ matrix.

\item \textbf{Step 3:} Verifier then checks whether the formed triangles, by the \textit{PPI}  of \textit{sweetwords} in $i^{th}$ round, are getting overlapped or not.

\item \textbf{Step 4:} If there is no overlapping,  $\mathcal{A}^+_\text{S3PAS}$ moves to \textit{Step 5} otherwise, goes back to \textit{Step 2}.

\item \textbf{Step 5:} Fixes the $m \times n$ matrix as obtained from \textit{Step 2} and displays the matrix on the visual interface.

\item \textbf{Step 6:} From the submitted response in each round, $\mathcal{A}^+_\text{S3PAS}$ checks that which triangles, formed by \textit{PPI} of \textit{sweetwords}, hold that response element. The verifier then records index position of the corresponding \textit{sweetwords}.

\item \textbf{Step 7:} At the end of all login rounds, verifier selects that \textit{sweetword}, whose index position is recorded in all the rounds.

\item \textbf{Step 8:} If no common index value is found then verifier denies the login otherwise, $\mathcal{A}^+_\text{S3PAS}$ moves to \textit{Step 9}. 

\item \textbf{Step 9:} Along with the username, index value is directed to  the \textit{honeyChecker} server.

\item \textbf{Step 10:} On the receiving the positive feedback from \textit{honeyChecker}, $\mathcal{A}^+_\text{S3PAS}$ allows login otherwise, blocks the user.

\end{itemize}

\begin{figure*}
\hfill
\subfigure[Formed triangles by the PPIes of first round.]{\includegraphics[width=4cm]{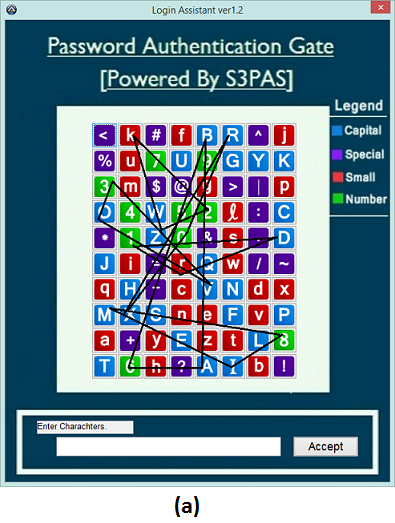}}
\hfill
\subfigure[Formed triangles by the PPIes of second round.]{\includegraphics[width=4cm]{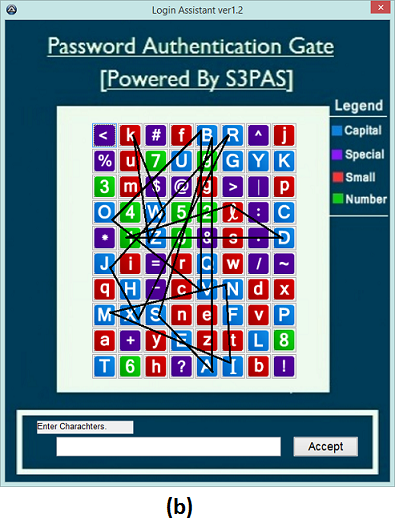}}
\hfill
\subfigure[Formed triangles by the PPIes of third round.]{\includegraphics[width=4cm]{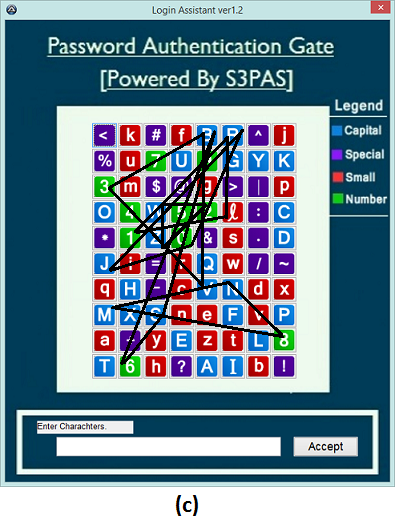}}
\hfill
\subfigure[Formed triangles by the PPIes of fourth round.]{\includegraphics[width=4cm]{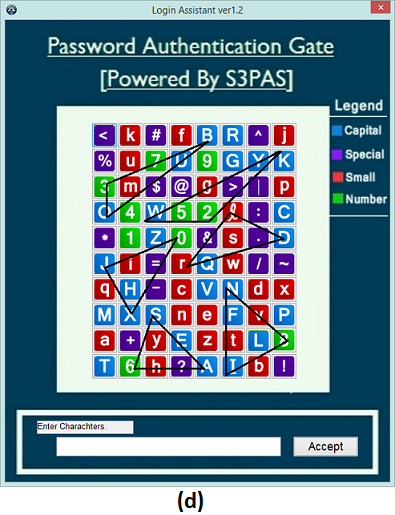}}
\hfill
\caption{Above figure presents visual interface returned by $\mathcal{A}^+_\text{S3PAS}$. This interface ensures that generated triangles in the fourth round, uniquely identify user response for the \textit{sweetwords} 2KZW, 8IMN, 6ABS, 0XRJ, 3OVB and rD1$\ell$.  \textbf{(a)} Formed overlapping virtual triangles in first round on the interface of modified S3PAS. Corresponding  \textit{PPI} of the \textit{sweetwords} are 2KZ, 8IM, 6AB, 0XR, 3OV and rD1. \textbf{(b)} Formed overlapping virtual triangles in second round on the interface of modified S3PAS. Corresponding  \textit{PPI} of the \textit{sweetwords} are KZW, IMN, ABS, XRJ, OVB and D1$\ell$. \textbf{(c)} Formed overlapping virtual triangles in third round on the interface of modified S3PAS. Corresponding  \textit{PPI} of the \textit{sweetwords} are ZW2, MN8, BS6, RJ0, VB3 and 1$\ell$r. \textbf{(d)} Formed non overlapping virtual triangles in fourth round on the interface of modified S3PAS. Corresponding  \textit{PPI} of the \textit{sweetwords} are W2K, N8I, S6A, J0X, B3O and $\ell$rd. }
\label{improved-s3pas}
\end{figure*}

We explain the above procedure with an example. Let for \textit{k} as $6$, a verifier maintains the following list of \textit{sweetwords} $-$ \textit{2KZW}, \textit{8IMN, 6ABS, 0XRJ, 3OVB} and \textit{rD1$\ell$}, in $F_{P}$.
Now suppose $\mathcal{A}^+_\text{S3PAS}$ randomly selects $4^{th}$ round to distinguish the original password from the \textit{honeywords}. In Figure \ref{improved-s3pas}, we show the visual interface (or challenge) generated by the verifier. The visual interface ensures that formed triangles in $4^{th}$ round share no common elements.

As shown in Figure \ref{improved-s3pas}, submitted response in any other round (except $4^{th}$) may belong to more than one triangles as they may share some common areas. But, atleast from the submitted response in the $4^{th}$ round, system will able to understand for which \textit{sweetword}, the response is being submitted. Thus, by following \textit{Principle $1$} and \textit{Principle $2$}, modified S3PAS can differentiate the original password from the set of \textit{honeywords}.

\subsubsection{Determining the value of \textit{k}}
\label{val-k}
According to \textit{Principle $3$}, the number of \textit{sweetwords} in $\mathcal{A}^+_\text{S3PAS}$ may vary between $2$ to $80$ (number of response elements). But as a \textit{PRS} generated by $\mathcal{A}^+_\text{S3PAS}$ is expected to hold more than one element, thus maximum limit of $80$ cannot be reached in practice. 
 
 $\mathcal{A}^+_\text{S3PAS}$ gets stuck in a loop (between \textit{Step 2} and \textit{Step 4}) until $k$ non-overlapping triangles are formed in a particular round, decided at \textit{Step 1}. The triangles need to be fitted on the visual interface which is nothing but a  $10 \times 8$ matrix of \textit{T} ($= 80$) elements. As \textit{k} non-overlapping  triangles are needed to be constructed on a fixed area, thus we make the following claim to determine a suitable value of \textit{k}.\\

\begin{theorem}
On a fixed area, if we randomly draw \textit{k} ($> 1$) triangles, then probability of overlapping among them is directly proportional to \textit{k}.
\end{theorem}

\textit{\textbf{Proof :}} Let us assume that the size of the visual interface is \textit{AR} units. The average area of each triangle is considered as \textit{TR} ($<$ \textit{AR}) units. The triangles can only be overlapped if there exists atleast one triangle prior to choosing the vertices for the second triangle. Thus, according to \textit{Roulette Wheel} selection principle \citep{roulette-wheel}, the probability of overlapping between \textit{k} $= 2$ triangles can be determined as $\text{TR}/\text{AR}$. Similarly, the probability that \textit{k}$^{th}$ triangle will overlap with any of the existing  \textit{k-1} triangles can be derived by using the following Equation \ref{k-val}.

\begin{equation}
(k-1)\times \dfrac{\text{TR}}{\text{AR}} \quad \text{where:} \quad k > 1 \quad and \quad k \times TR \leq AR
\label{k-val}
\end{equation}

From the above equation, it is easily understandable that for drawing \textit{k} non-overlapping triangles on a fixed area, the value of \textit{k} can not be arbitrarily large.\\

\textbf{Experimental analysis for determining the value of k:} We also perform an experimental analysis for choosing a suitable value of \textit{k}. In $\mathcal{A}^+_\text{S3PAS}$, there is a loop from \textit{Step 2} to \textit{Step 4}. The loop only breaks if \textit{k} non-overlapping triangles are generated in a around, predetermined by the $\mathcal{A}^+_\text{S3PAS}$. From Equation \ref{k-val}, it is quite evident that with the increasing value of \textit{k}, formation of non-overlapping triangles on a fixed plane will be difficult. Thus, we determine a suitable value of \textit{k} based on the average number of iterations and average time consumed by the loop. 

\begin{table*}[!ht]
\centering
\resizebox{1\textwidth}{!}{
\begin{tabular}{|c|c|c|c|c|c|c|}
\hline
\textbf{value of k} & \textbf{no. of max iteration} & \textbf{no. of min iteration} & \textbf{avg iteration} & \textbf{max exec time (ms)} & \textbf{min exec time (ms)} & \textbf{avg exec time (ms)}\\
\hline
4 & 34 & 1 & 10 & 1132 & 1018 & 1045 \\
\hline
5 & 48 & 1 & 16 & 1246 & 1012 & 1083 \\
\hline
6 & 251 & 3 & 63 & 2530 & 1032 & 1391 \\
\hline
7 & 995 & 9 & 402 & 7088 & 1068 & 3712 \\
\hline
8 & 4043 & 174 & 2146 & 28791 & 2628 & 16205\\
\hline
\end{tabular}}
\caption{Above table shows for different values of \textit{k}, number of iterations and running time required by the \textit{loop} to generate \textit{k} non-overlapping triangles in a round. The data is collected after $20$ runs of $\mathcal{A}^+_\text{S3PAS}$ for each value of \textit{k}.}
\label{run-time}
\end{table*}

Table \ref{run-time} reflects the details of experimental analysis performed on a system having software specification $-$ OS: \textit{Windows $7$}, PHP version: \textit{$5.4.3$} and Server API: \textit{Apache $2.0$ Handler} and, hardware specification $-$ processor: \textit{$32$ bit core $i3$} and primary memory: \textit{$4$GB}. To conduct the experiment, we have varied \textit{k} from $4$ to $8$.  
From Table \ref{run-time}, it can be seen that for \textit{k} as $7$, the system took $7.08$ seconds at max (on average $3.7$ seconds) to produce a challenge which can differentiate among $7$ \textit{sweetwords} from the response of a user. When  \textit{k} reached to $8$, the average time for generating visual interface was $16.2$ seconds. Thus, while $k$ was set to $7$ and $8$, we found that it affected the session (login) time and usability standard. So, we prefer the value of \textit{k} as $6$ over the others as it took  approx $1.39$ seconds on average (and that of $2.53$ seconds at max) to generate the visual interface.  \\

Also, as large value of $k$ covers more response elements, thus selecting a higher value of $k$ may degrade the security standard of the modified scheme against DoS attack (see Section \ref{hsf}). Therefore, \textit{k} as $6$ does not put any significant overhead on the session time and, with moderate security against DoS (also see details in Section \ref{sec-ana-s3pas}), it provides a healthy detection rate of $83.34\%$.

\subsubsection{Security analysis of modified S3PAS}
\label{sec-ana-s3pas}
As mentioned in Section \ref{msp}, modification from $\mathcal{A}_\text{S3PAS}$ to $\mathcal{A}^+_\text{S3PAS}$   may have some impacts on the existing security standard of S3PAS. Also security notion of HBAT framework gets added to the basic security standard to reinforce the overall security level. To start with, we show how proposed modification influences the basic security features of S3PAS. \\

\textbf{Basic security features:} The basic security feature of S3PAS includes security against \textit{observation attack through brute force search}, \textit{password guessing attack} and \textit{random click attack} as shown in \citep{strong-ssa-s3pas}. Thus, we determine the basic security standard of modified S3PAS against these three attack models.\\

\textbf{(a) Security against observation attack through brute force search:} Brute force attack is a general pruning-based learning process, where the eavesdropper keeps eliminating irrelevant candidates when more and more cues are available. To perform this attack, the adversary first lists all possible candidates for the password and for each independent observation of challenge response pair, she checks the validity of each password candidate by running the verification algorithm used by the server. Thereafter she discards invalid candidates from the candidate set. The aforementioned procedure is continued until the attacker derives a candidate set of small threshold.

The complexity of this attack for basic S3PAS scheme can be derived by following the same path as proposed in \citep{CHC} and can be represented in the form of  \large O(\normalsize $|T| \choose |\text{PPI}|$\large)\normalsize; where $|$PPI$|$ denotes the length of \textit{PPI} whose default value is $3$. As values of both $|$T$|$ and $|$PPI$|$ remain unchanged in modified S3PAS, thus security standard against this attack remains same for both the variations of S3PAS.\\

\textbf{(b) Security against password guessing attack:} To perform this attack, an adversary tries to guess actual password to pass through a session \citep{password-guess-attack}. Therefore, success probability of adversary depends upon two components $-$ length of the password and the password space. As values of both these components remain same for both the existing and modified schemes, thus modified S3PAS maintains same security standard against this threat.\\

\textbf{(c) Security against random click attack:} Security against this attack is heavily dependent on the expected triangle area (denoted as E[S]) \citep{strong-ssa-s3pas}. Except in $i^{th}$ round (dedicated to form $k$ non-overlapping triangles on the user interface), modified scheme does not impose any restriction on the area of the triangles in any other round. In \citep{strong-ssa-s3pas}, authors determine the E(S) in the from of the following equation

\begin{equation}
\text{E[S]} = \sum_{f=1}^{n}\sum_{g=1}^{n}\sum_{h=1}^{n}\sum_{i=1}^{n}\sum_{j=1}^{n}\sum_{k=1}^{n}\dfrac{1}{2}|(\dfrac{f}{n}-\dfrac{g}{n})(\dfrac{i}{n}-\dfrac{k}{n})-(\dfrac{f}{n}-\dfrac{h}{n})(\dfrac{i}{n}-\dfrac{j}{n})|\dfrac{1}{n^6}
\end{equation}
 
For the basic S3PAS scheme, we may consider value of $n$ as $9$ here as a grid of dimension $9 \times 9$ is capable of holding all \textit{T} $= 80$ elements. Therefore for $n = 9$ the above equation yields to $0.753$. This infers that probability of success under random click attack in each round is $0.753$.

In the modified scheme, as value of \textit{T} remains same as of the basic S3PAS, thus proposed modification provides same security standard against this attack too.\\

\textbf{Inherited HBAT security features:} As discussed earlier, HBAT security feature is comprising of security against DoS and MSV. Achieving flatness is another security property which comes under HBAT framework. Next we show how well modified S3PAS satisfies these HBAT security features.\\

\textbf{(a) Security against DoS attack:}  As discussed in Section \ref{hsf}, to provide robust security against DoS, a method belonging to $\text{M}^\text{FODS}_\text{SOA}$ must satisfy following two criteria $-$
\begin{itemize}
\item Generated \textit{honeywords} must be hard to guess.
\item Value of $k$ should be chosen in such a manner so that it does not cover all the response elements. 
\end{itemize}

For generating hard to guess \textit{honeyword}, modified S3PAS may use modelling-syntax-approach \citep{kamouflage}. As value of $k$ cannot be chosen arbitrarily large (ref. to Section \ref{val-k}), therefore by virtue of its design, proposed modification satisfies second criterion too. Thus we may claim that modified S3PAS successfully mitigates the threat of DoS attack.\\

\textbf{(b) Security against MSV:} Though modelling-syntax-approach provides good security to resist DoS attack, but it provides weak security against MSV. In fact, till date there exists no solution which provides strong security against both these attacks without having any system interference. System interference degrades the usability standard of a HBAT to a great extent.\\

\textbf{(c) Flatness:} Modelling-syntax-approach can generate absolutely flat list of \textit{sweetwords}. Thus, modified S3PAS achieves absolute flatness, only if there exists no correlation between the username and password.\\
 
Therefore, having some scope of improvements in terms of providing security against MSV, modified S3PAS satisfies almost all security parameters.

\subsubsection{Usability analysis of modified S3PAS}
\label{u-s3pas}
As mentioned in Section \ref{mup}, to get compatible with HBAT framework,  modification of $\mathcal{A}_\text{S3PAS}$ may influence the login behaviour of a user. Moreover, usability features of HBAT are also added to determine the overall usability standard. \\

\textbf{Basic usability features:} The usability standard of the existing scheme can be determined with respect to two parameters $-$ \textit{login time} and, \textit{percentage of error} during login. Modified scheme imposes no \textit{CSLI} and the login procedure remains exactly same as of the existing one. A user even does not require to remember any extra information to get compatible with the proposed architecture. Thus, proposed scheme can be categorized as $-$ no \textit{CSLI} with no \textit{CLP} and no \textit{REI} (ref. to Section \ref{mup}) and hence, modified S3PAS provides same usability standard as of the existing one.\\

\textbf{Inherited HBAT usability features:} Modified S3PAS makes no impact on the password choice of the user and hence, it neither imposes any stress on user's mind nor has any system interference. Typing mistake of a genuine user will set off an alarm by the \textit{honeyChecker} server, only if, response submitted by the user in each round corresponds to \textit{PPI} of a particular \textit{honeyword}. Let each character takes $1$ unit area to fit into the visual interface of S3PAS. Let the average area of each triangle be \textit{TR} units, then for \textit{T} ($= 80$) response elements, probability of submitting a response element belonging to a  \textit{PRS} (formed by the \textit{PPI}) of a particular \textit{honeyword} becomes $\frac{\text{TR}}{\text{T}}$. Thus for an authentication session of \textit{lr} ($= 4$) login rounds, probability of hitting the same \textit{honeyword} becomes $(\frac{\text{TR}}{\text{T}})^{lr}$. While determining  a suitable value of \textit{k} (see Section \ref{val-k}), we found that for the value $k = 6$, a virtual triangle mostly holds $3$ elements in it. Therefore by considering \textit{TR} $= 3$, for the default values of parameters, the probability can be derived as $1.9 \times 10^{-6}$ . So we may claim that proposed approach is typo safe.\\

\textbf{Note 4:} Proposed modification of basic S3PAS allows no degradation of basic security and usability standards. We also show that except MSV, modified S3PAS satisfies all security parameters of HBAT with moderate value of $k = 6$ (particularly helpful to resist the DoS attack). Finally, modified scheme well supports all HBAT usability features to provide adequate user friendliness.

\subsection{Incorporating honeywords to CHC}
In this section, first we briefly introduce Convex-Hull-Click (CHC) protocol \citep{CHC} and then show how HBAT can be incorporated to this.

\subsubsection{Working principle of CHC} From their functional aspect, CHC  and S3PAS are very similar in nature \citep{yan-2015}. In CHC, instead of choosing a password of length $4$, user selects $\mathcal{K} (\geq 3)$ pass icons as her secret out of $\mathcal{N}$ icons. At each round, system randomly chooses $\mathcal{M}$ icons including $\mathcal{K}^{c}$ ($3 \leq \mathcal{K}^{c} \leq \mathcal{K}$) pass icons and display those on the login terminal. To give her response, a user first locates those $\mathcal{K}^{c}$ pass icons on the login screen and makes a convex hull by using those. Then the user selects a point (icon) inside that convex hull and submits that as a response.

It is very easy to relate that for basic S3PAS, $\mathcal{N} = \mathcal{M}$ and we select value of that as $80$. Also, at each round in S3PAS, the user needs to locate $\mathcal{K}^{c} = 3$ consecutive characters from her password of length $ \mathcal{K} = 4$.

\subsubsection{Proposed modification for incorporating HBAT framework}
Due to similarity between the approaches, proposed modification here follows the same path as of S3PAS. System should choose $k$ \textit{sweetwords} in such a manner so that at any given round, \textit{PPI} of those are different (ref. to \textit{Principle} $2$). To authenticate the user in a session, system first randomly selects a round and ensures that with reference to the visual challenge in that particular round, generated $k$ convex hulls by the \textit{PPI} of the \textit{sweetwords} share no common area (ref. to \textit{Principle} $1$). Thus at the end of the authentication session, from the submitted response by the user, system can uniquely identify the \textit{sweetword} against which the login has been performed. System then follows the normal HBAT routine to detect the breach.  

\subsubsection{Security analysis of modified CHC}
Along with basic security characteristics, Security analysis here includes inherited security features due to incorporation of HBAT. As shown in \citep{CHC}, the basic security feature comprises of security against \textit{observation attack through brute force search} and \textit{password guessing attack}. Also, due to difference in distribution of secret icons and non secret icons, security of basic CHC may be threatened under \textit{probabilistic} attack too [\citep{cryptCHC} Section $4$, pp. 6]. For modified CHC, next we elaborate these attack scenarios one by one.\\

\textbf{(a) Security against observation attack through brute force search:} Security analysis against this attack is done in the same way as proposed in \citep{CHC}. Attacker initially forms a list containing all possible $\mathcal{K}$ combinations of icons out of $\mathcal{N}$ icons. After recording each challenge response pair, attacker then discards all those combinations from the list whose convex hull do not include that response point. The sole remaining $\mathcal{K}$ combination is then become the $\mathcal{K}$ secret labels of the user. Therefore, order the attack can be derived as \large O(\normalsize $\mathcal{N} \choose \mathcal{K}$\large)\normalsize. As modified scheme does not influence the parameters of existing CHC protocol, therefore, the complexity of the attack remains same for the proposed modification. \\

\textbf{(b) Security against password guessing attack:} Using the modified approach, user is allowed to select the same password as of the existing one. Hence, security standard against password guessing attack does not get changed.\\

\textbf{(c) Security against probabilistic attack:} In CHC, in each round, out of $\mathcal{N}$ icons $\mathcal{M}$ icons are displayed on the visual interface and, visual interface always contains $\mathcal{K}^{c}$ ($3 \leq \mathcal{K}^{c} \leq \mathcal{K}$) pass icons. But there is no such restriction on non pass icons. For the basic CHC scheme, authors in \citep{cryptCHC} present the expected number of times an icon (\textit{I}) appears in $r$ challenges in the form of following equation

\begin{equation}
\text{E[}I^r, I \not\in \mathcal{K}\text{]} = r \times \dfrac{1}{\mathcal{K}-2} \times \dfrac{1}{\mathcal{N}-\mathcal{K}} (\mathcal{M}(\mathcal{K}-2) - \dfrac{\mathcal{K} (\mathcal{K}+1)}{2} +3)
\label{chc-eq-1}
\end{equation} 
where \textit{I} does not belong to the set of pass icons.\\

and,

\begin{equation}
\text{E[}I^r, I \in \mathcal{K}\text{]} = r \times \dfrac{1}{\mathcal{K}(\mathcal{K}-2)} \large(\normalsize \dfrac{\mathcal{K} (\mathcal{K}+1)}{2} -3\large)\normalsize
\label{chc-eq-2}
\end{equation}

where \textit{I} belongs to the set of pass icons.\\

By following Equation \ref{chc-eq-1} and Equation \ref{chc-eq-2}, in \citep{cryptCHC}, authors derive that for $\mathcal{N} = 112$, $\mathcal{M} = 70$, $\mathcal{K} = 5$ and $r = 100$,  secret icons and non secret icons are expected to appear $80$ times and $62$ times, respectively.  Thus, in basic CHC, observation of $\mathcal{K}$ most frequently occurring icons in $r$ challenge help adversary to get the secret.

In the modified CHC, though values of $\mathcal{N}$ and $\mathcal{M}$ remain same, but due to incorporating \textit{honeywords}, number of \textit{sweet icons} increases. This in turn increase number of icons that must appear on visual interface in a round. We denote number of \textit{sweet icons} as \textit{I}$^{S}$; where \textit{I}$^S$ $> \mathcal{K}$. While replacing $\mathcal{K}$ in the above equations by \textit{I}$^S$, we always expect to get a lesser value from Equation \ref{chc-eq-2}, and thus, we may claim that to obtain all \textit{sweet icons}, an attacker needs to go through more number of rounds. \\

\textbf{Inherited HBAT security features:} As modified CHC inherits the HBAT in almost the same way as of modified S3PAS, therefore it can provide good security in terms of achieving flatness and defending DoS. But this method falls short while building security against MSV. 

\subsubsection{Usability analysis of modified CHC}
Usability analysis of modified CHC protocol comprises of basic usability features and usability parameters of HBAT. As user remembers same password information and does not face any change during login, therefore without any \textit{REI} from user end, modified \textit{CHC} imposes no \textit{CSLI} and no \textit{CLP}. This infers that basic usability standard remains same for both the existing CHC and modified one.

Also from HBAT usability aspect too, this method can be considered as typo safe (because of the similar reason discussed in Section \ref{u-s3pas}) without any system interference and no stress-on-memorability.\\

\textbf{Note 5:} Likewise modified S3PAS, proposed modification of CHC maintains the same basic security standard in terms of resisting brute force attack and password guessing. In fact, modified CHC performs better by increasing the complexity of probabilistic attack. The basic usability standard is maintained same as of the existing CHC protocol. Also, except MSV, modified CHC satisfies all security and usability aspects of HBAT.

\subsection{Incorporating honeywords to PAS}
Under the scope of this section, first, we briefly describe the basic PAS scheme \citep{strong-ssa-pas} and, thereafter we show a direction towards integrating HBAT into this.

\subsubsection{Working principle of PAS}
To login into the system, a user remembers two predicates as secret. Each predicate contains one index value along with an alphabet. For example, \textit{23E} and \textit{41P} are two valid predicates chosen by the user. The login interface of PAS contains two tables. Each table is made of $25$ blocks and each block is denoted by an index value starting from  $(1,1)$ to $(5,5)$. Along with a block number (i.e., an index value), each block holds $13$ alphabets chosen randomly from the set of $26$ alphabets. The first cell of a block always contains the index value.

\begin{figure*}[!ht]
\centering
\includegraphics[width=0.8\textwidth]{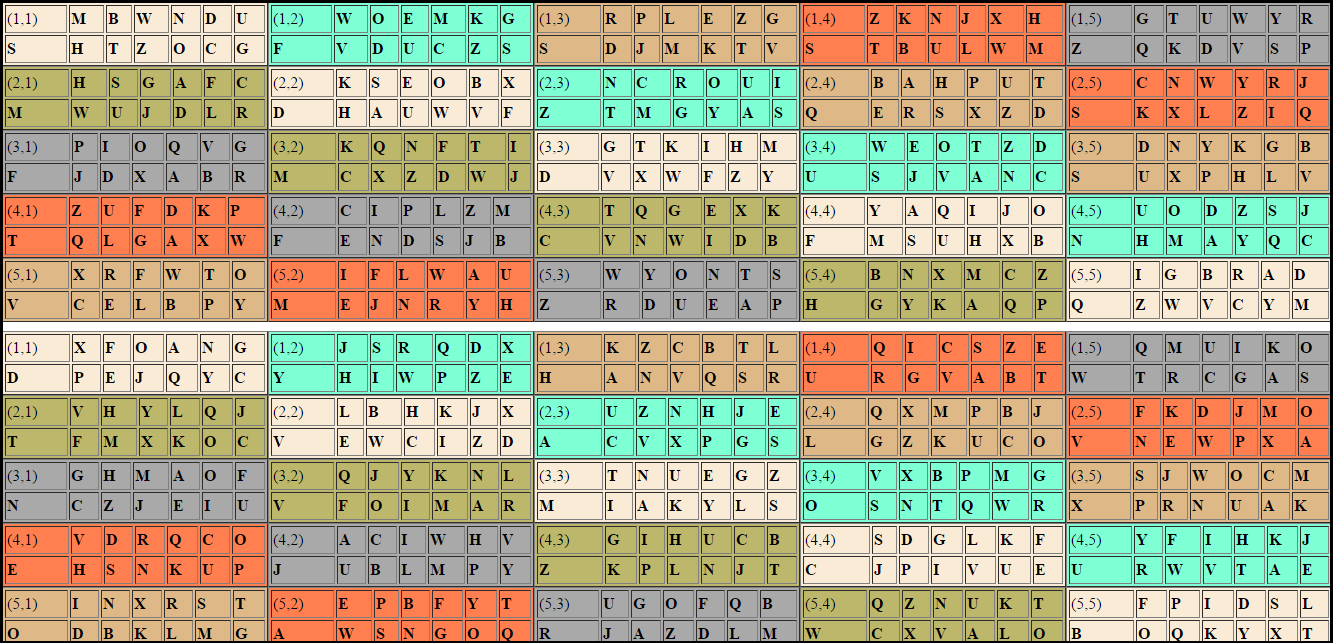}
\caption{Challenge tables in PAS}
\label{c-tab-pas}
\end{figure*}   

In Figure \ref{c-tab-pas}, we have shown two challenge tables of PAS. To enter response with respect to her first predicate (i.e., \textit{23E}) and the challenge shown in Figure \ref{c-tab-pas}, the user first moves to the block denoted by (2, 3) in the first challenge table and searches for character \textit{E} there. 
As \textit{E} is not present in that block, therefore user derives \textit{NO} as answer. For searching \textit{E} again, the user then moves to the block number (2, 3) in the second challenge table. As block (2, 3) contains character \textit{E} in this table, thus user derives \textit{YES} as her answer. Therefore, from the first predicate \textit{23E} and with respect to the challenge tables shown in Figure \ref{c-tab-pas}, the user derives the answer sequence \textit{NO, YES}.

\begin{figure}
\centering
\includegraphics[width=0.5\textwidth]{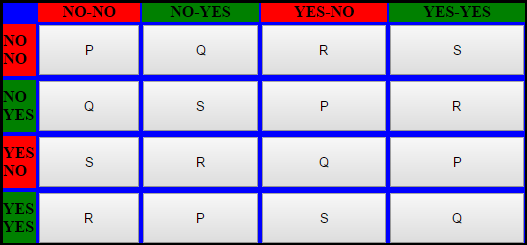}
\caption{Response table in PAS}
\label{r-tab-pas}
\end{figure}

In the same way, with respect to her second predicate (i.e., \textit{41P}) and the challenge tables shown in Figure \ref{c-tab-pas}, derived answer sequence by the user will be \textit{YES, YES}. User remembers this cumulative answer sequence as $\{$NO, YES, YES, YES$\}$. To give a response with respect to the derived answer sequence, the user takes help of the response table shown in Figure \ref{r-tab-pas}.

The derived answer from the first predicate and the second one correspond to the row index and the column index of the response table, respectively. Thus, user derives her response as \textit{R} with reference to this example. After typing the response user is led to the next round. It is important to note here that in the remaining login rounds, with changes in the content of challenge tables, the user follows the same  procedure to generate the responses with respect to the same predicate pair \textit{23E} and \textit{41P}. 

\subsubsection{Proposed modification for incorporating HBAT framework}
In the basic PAS scheme, user chooses her response from a set of $4$ elements $\{$P, Q, R, S$\}$. Therefore, first of all, according to the \textit{Principle} $3$, the value of $k$ cannot go beyond $4$. Next, as \textit{PPI} of \textit{sweetwords} are required to be different (ref. to \textit{Principle} $2$) and predicates act as \textit{PPI} here, thus modified \textit{PAS} needs to generate predicates of \textit{honeyword} in such a manner so that they are unique. To generate different predicates, system may either vary the index in a predicate, or the character in it, or both. With out loss of generality, our proposal here alters both to generate the \textit{honeywords} with respect to the original predicates of the user.  

In PAS, user uses two predicates to pass an entire authentication session. Hence, to satisfy \textit{Principle} $1$, modified PAS should generate the challenge in such a manner so that atleast in a round, predicates of each \textit{sweetword} map to a unique response set. It is important to note here that in PAS, each response set holds a single element only. Now as shown in Figure \ref{r-tab-pas}, each response in the response table gets mapped to $4$ answer sequences and, the relation between response elements and answer sequences is shown in Table \ref{map-element}.

\begin{table}[!ht]
\centering
\resizebox{0.95\textwidth}{!}{
\begin{tabular}{|c|c|}
\hline
Response element & Corresponding answer sequences\\
\hline
P                & $\{$NO, NO, NO, NO$\}$, $\{$NO, YES, YES, NO$\}$, $\{$YES, NO, YES, YES$\}$, $\{$YES, YES, NO, YES$\}$\\
\hline
Q                & $\{$NO, NO, NO, YES$\}$, $\{$NO, YES, NO, NO$\}$, $\{$YES, NO, YES, NO$\}$, $\{$YES, YES, YES, YES$\}$\\
\hline
R                & $\{$NO, NO, YES, NO$\}$, $\{$NO, YES, YES, YES$\}$, $\{$YES, NO, NO, YES$\}$, $\{$YES, YES, NO, NO$\}$\\
\hline
S                & $\{$NO, NO, YES, YES$\}$, $\{$NO, YES, NO, YES$\}$, $\{$YES, NO, NO, NO$\}$, $\{$YES, YES, YES, NO$\}$\\
\hline
\end{tabular}}
\caption{Relationship between the response elements and the answer sequences with reference to the Figure \ref{r-tab-pas}.}
\label{map-element}
\end{table}

To satisfy \textit{Principle} $1$ in $i^{th}$ round ($1 \leq i \leq lr$), modified PAS may adopt the following strategy. In round $i$, it first randomly selects one answer sequence corresponding to each response. Then one of these (here $4$) selected answer sequences is assigned to each \textit{sweetword}. During formation of challenge tables, each \textit{sweetword} then try to satisfy the answer sequence assigned to it. Satisfying a answer sequence automatically satisfies corresponding response element. Thus, each \textit{sweetword} maps to a different response through the communicated challenge in the $i^{th}$ round. Except in $i^{th}$ round, multiple \textit{sweetwords} may map to a single response element in other rounds. \\ 

For example, with respect to the original predicates as ($23$\textit{E}, $41$\textit{P}) system selects other $3$ predicate pairs as ($32$\textit{S}, $51$\textit{T}), ($34$\textit{Y}, $11$\textit{M}) and ($15$\textit{Z}, $55$\textit{B}) and hence satisfies both the principles $-$ \textit{Principle} $2$ and \textit{Principle} $3$.\\ 

\begin{figure*}[!ht]
\centering
\includegraphics[width=0.8\textwidth]{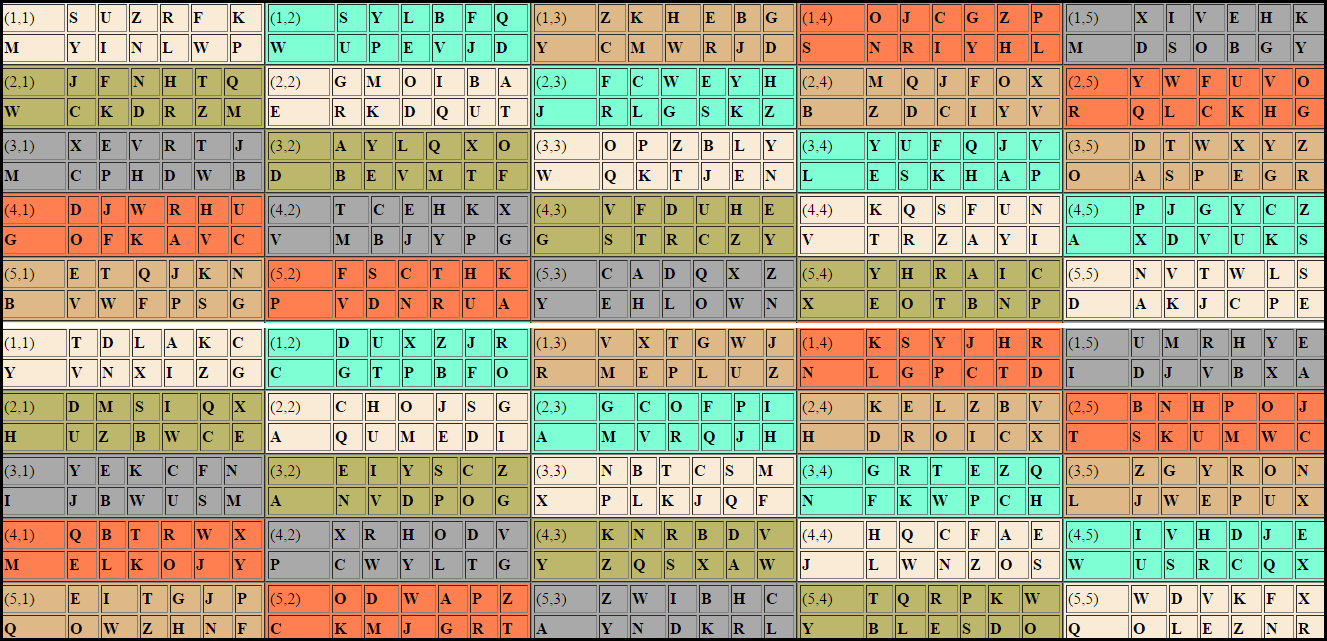}
\caption{Challenge table in modified PAS in $i^{th}$ round}
\label{m-tab-pas}
\end{figure*}

Let the assigned answer sequence to the original predicates be $\{$YES, NO, NO, NO$\}$ which satisfies the response element \textit{S}. Hence, while first predicate (i.e., \textit{23E}) ensures that alphabet \textit{E} appears in block (2, 3) in the first table and does not appear in the same block of the second table, the other predicate (i.e., \textit{41P}) restricts alphabet \textit{P} from appearing in block (4, 1) in both the tables. In such manner, predicate ($23$\textit{E}, $41$\textit{P}) satisfies the answer sequence $\{$YES, NO, NO, NO$\}$ and maps to the response element \textit{S}.  In Figure \ref{m-tab-pas}, we show the challenge interface in which (with reference to the response table shown in Figure \ref{r-tab-pas})
\begin{itemize}
\item Predicate ($15$\textit{Z}, $55$\textit{B}) is satisfying answer sequence $\{$NO, NO, NO, NO$\}$ and maps to the response element \textit{P}.
\item Predicate ($34$\textit{Y}, $11$\textit{M}) is satisfying answer sequence $\{$ YES, NO, YES, NO$\}$ and maps to the response element \textit{Q}.
\item Predicate ($32$\textit{S}, $51$\textit{T}) is satisfying answer sequence $\{$NO, YES, YES, YES$\}$ and maps to the response element \textit{R}.   
\item Predicate ($23$\textit{E}, $41$\textit{P}) is satisfying answer sequence $\{$YES, NO, NO,  NO$\}$ and maps to the response element \textit{S}.
\end{itemize}

Therefore, using proposed modification of PAS, even though there may exist some overlapping in other rounds, but from the submitted response in the $i^{th}$ round, system can uniquely derive the \textit{sweetword} against which user performs the login operation.  For generating the \textit{honeywords}, likewise modified S3PAS and modified CHC protocols, modelling-syntax-approach \citep{kamouflage} may also be adopted here.

\subsubsection{Security analysis of modified PAS}
Security analysis of modified PAS scheme includes basic security features and inherited HBAT security features. To start with we will focus on basic security features.\\

\textbf{Basic security features:} As shown in \citep{sec-PAS}, the basic security features of PAS include security against \textit{observation attack through brute force search} and \textit{password guessing attack}. Next we elaborate these one by one.\\

\textbf{(a) Security against observation attack through brute force search:} Under this attack model, an attacker initially considers all possible  predicate pair and narrows the list after observing each challenge response pair. In \citep{sec-PAS}, authors show that computational complexity of this attack can be determined as \large O(\normalsize ${MH+c-1 \choose c}^p$ \large) \normalsize where \textit{M}, \textit{H}, \textit{c} and \textit{p} denote number of cells in the challenge table, number of all possible characters, number of indices in each predicate and number of predicates in the secret, respectively. As modified  PAS does not make any impact on these parameters, therefore, it maintains the same security standard against this attack as of the basic PAS scheme.\\

\textbf{(b) Security against password guessing attack:} Success probability of an attacker under this attack scenario depends on the password space and the length of the password. As both these factors remain unchanged in the modified scheme, thus basic PAS and modified one provide same security standard against this attack model too.\\

\textbf{Inherited HBAT security features:} As discussed previously, along with achieving flatness, integrated HBAT security feature includes security against DoS attack and MSV.\\

\textbf{(a) Security against DoS attack:} As user response is confined within $4$ response elements only, thus, mounting DoS attack becomes easy for the proposed modified PAS. Therefore, as suggested in \citep{honeyword-erguler}, modified PAS may adopt light security policy against DoS attack.\\

\textbf{(b) Achieving flatness and security against MSV:} As modelling-syntax-approach can be used for generating the \textit{honeywords}, therefore, though proposed scheme achieves complete flatness, but cannot provide robust security against MSV.

\subsubsection{Usability analysis of modified PAS}
Usability analysis of proposed modification is done against basic usability features and integrated HBAT usability features.\\

\textbf{Basic usability standard:} As modified PAS falls under the category of no \textit{CSLI} with no \textit{CLP} and no \textit{REI}, thus it provides the same usability standard as of the basic PAS scheme.\\

\textbf{Inherited HBAT usability standard:} Proposed modification here neither imposes stress on user mind nor have any system interference. Probability of typing mistake, that raises the false alarm, is influenced by two factors $-$ number of login rounds (\textit{lr} $= 5$) and probability of submitting the wrong response (here $3/4$), and can be derived as $\frac{3}{20}$ for the default values of parameters. Thus with moderate typo-safety, modified PAS well satisfies all other HBAT usability parameters. \\

\textbf{Note 6:} Modified PAS does not affect the basic security and usability standards of the basic PAS scheme. From HBAT security point of view, it provides weak security against DoS which can be handled by adopting light security policy against this attack as suggested in \citep{honeyword-erguler}. Proposed modification here though cannot robustly handle the MSV crisis, but achieves absolute flatness to obfuscate the attacker properly. Form the HBAT usability aspect too, modified scheme well satisfies almost all the usability parameters.

\subsection{Integrating honeywords to Count-On-Plane}
\label{cop}
In this section, first we briefly introduce the proposed protocol in \citep{acns} which we name as Count-On-Plane (COP) in this literature. Thereafter, we show how HBAT features can be added to this protocol by following the proposed principles in Section \ref{p-c-h}.

\subsubsection{Working principle of COP}
\begin{figure}[!ht]
\centering
\includegraphics[width=0.6\textwidth]{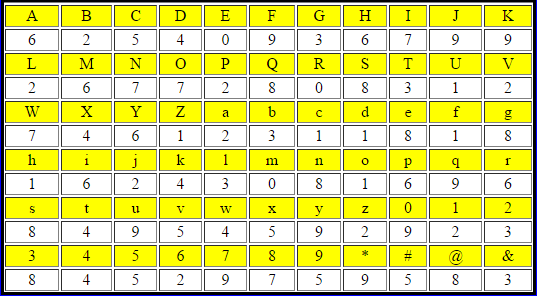}
\caption{Challenge grid in COP}
\label{cop-interface}
\end{figure}

To login through COP, a user remembers a secret of length $\ell$ (here $4$) from a set of $66$ characters comprising of $\{$A,..., Z, a,...,z, 0,..., 9, $*$, $\#$, $@$, $\&$ $\}$. By maintaining a specific order, these characters are arranged on a 2D plane of dimension $n = a \times b$. On the 2D plane (visual interface), a value from $0$ to $9$ is randomly assigned to each character of the set. In Figure \ref{cop-interface}, we have shown an instance of challenge interface of COP protocol for $a = 11$ and $b = 6$. 
Yellow cells in the grid hold the characters from the set of $66$ elements and their position remain static in each session. White cells in the grid contain a random number from $0$ to $9$ and content of these cells change in every session.

Let the password remembered by a user be \textit{A1B3}. To login (with reference to the challenge grid shown in Figure \ref{cop-interface}), the user first goes to the character cell \textit{A} and looks at the digit corresponding to it. As user finds the digit as $6$ here, therefore in a circular way, she moves $6$ steps vertically downwards, and  thus reaching the same character location, \textit{A}. Except first, for the remaining characters of her password (i.e., \textit{1B3}), user then adds all the digits assigned to these characters. For this example, for the remaining characters $1B3$, this yields to $2 + 2 + 8 = 12$. From character position \textit{A} (reached after vertical shifts), the user then moves $12$ steps horizontally to reach to character \textit{M} and outputs the digit (here $6$) corresponding to \textit{M}. 

\subsubsection{Proposed modification for incorporating HBAT framework}
In COP protocol, as a user chooses her response from the set of $10$ elements $\{$0,...,9$\}$, therefore the value of $k$ can be reached upto $10$ at maximum (ref. to \textit{Principle} $3$). For providing better security against DoS, we consider value of $k$ as $5$ here. Next, to satisfy \textit{Principle} $2$, modified COP must generate $k$ ($1 \leq k \leq 10$) \textit{sweetwords} in such a manner so that they are different and also satisfy \textit{Principle} $1$.

To meet with \textit{Principle} $2$, modified COP produces $k$ ($= 5$) \textit{sweetwords} in such a manner so that they share no common element among them. For example, for original password as \textit{A1B3}, system may generate $4$ other \textit{honeywords} as $-$ \textit{QJw9}, \textit{2XTD}, \textit{YSRK} and \textit{icat}. To give her response, as a user must reach to  a particular cell (identified as response cell) and enter the digit corresponding to it, therefore, to validate a user under the light of integrated HBAT, we adopt the following strategy.

\begin{itemize}
\item \textbf{Step 1:} For length of each \textit{sweetword} as $\ell = 4$, except the elements of \textit{sweetwords}, system first selects $k = 5$ response cells randomly from $66 - (4 \times 5) = 46$ probable response cells.

\item \textbf{Step 2:} System them assigns a unique value (between $0$ to $9$) to each of the selected response cells.

\item \textbf{Step 3:} System then assigns each response cell to a \textit{sweetword}.

\item \textbf{Step 4:} For each \textit{sweetword}, system then does the following.

\begin{itemize}
\item \textbf{Step 4.1:} In a circular way, system first calculates the path length between the first element of a \textit{sweetword} and the corresponding response cell assigned to it. Let the derived path length be \textit{P}$_\text{L}$.

\item \textbf{Step 4.2:} \textit{P}$_\text{L}$ is then get divided by \textit{a}.

\item \textbf{Step 4.3:} System assigns the value of quotient to first element of the \textit{sweetword}.

\item \textbf{Step 4.4:} The remainder (r) gets partitioned  into $\ell - 1$ parts (say a, b and c) such that r = a + b + c. 

\item \textbf{Step 4.5:} The partition values are then assigned to the remaining characters of the \textit{sweetword}.
\end{itemize}

\item \textbf{Step 5:} This type of setup maps each \textit{sweetword} to a unique response cell, assigned by the system in \textit{Step 3}.

\item \textbf{Step 6:} As each response cell contains a unique value, thus it enables the system to understand that against which \textit{sweetword} the login has been performed. 
 
\end{itemize}  
From the above discussion, it is understandable that each \textit{sweetword} maps to a unique response set of cardinality $1$ to satisfy the \textit{Principle 1} and \textit{PPI} in COP are nothing but the whole password string.
Next, we discuss the above procedure with an example.\\

 For the \textit{sweetwords} as \textit{A1B3}, \textit{QJw9}, \textit{2XTD}, \textit{YSRK} and \textit{icat}, system randomly assigns  

\begin{itemize}
\item  response cell \textit{Z} to \textit{A1B3} having \textit{P}$_\text{L} = 25$ from A to Z. A value $3$ is assigned to \textit{Z}.
\item  response cell \textit{C} to \textit{QJw9} having \textit{P}$_\text{L} = 52$ from Q to C. A value $1$ is assigned to \textit{C}.
\item  response cell \textit{M} to \textit{2XTD} having \textit{P}$_\text{L} = 24$ from 2 to M. A value $5$ is assigned to \textit{M}.
\item  response cell \textit{H} to \textit{YSRK} having \textit{P}$_\text{L} = 49$ from Y to H. A value $6$ is assigned to \textit{H}.
\item  response cell \textit{h} to \textit{icat} having \textit{P}$_\text{L} = 65$ from i to h. A value $8$ is assigned to \textit{h}.
\end{itemize}

\begin{figure}[!ht]
\centering
\includegraphics[width=0.6\textwidth]{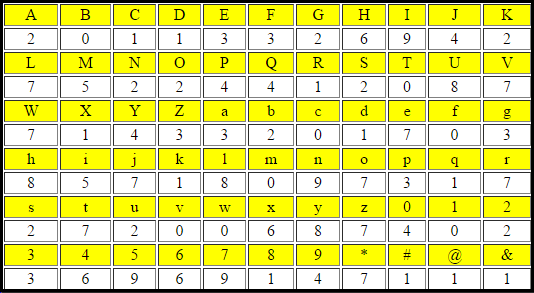}
\caption{Challenge grid in modified COP}
\label{modified-cop-interface}
\end{figure}

For \textit{a} $= 11$, system then divides $25$ by $11$ and assigns the quotient $2$ to \textit{A}. The remainder $25-22 = 3$ gets partitioned into $0, 0$ and $3$ and are allocated to $1$, \textit{B} and $3$, respectively (ref. to Figure \ref{modified-cop-interface}). System follows the aforementioned strategy for assigning values to the elements of all other \textit{sweetwords}.

Now, as challenge (see Figure\ref{modified-cop-interface}) is generated in such a way so that each \textit{sweetword} maps to a unique response element, therefore, from the response submitted by the user, system will uniquely be able to identify the corresponding \textit{sweetword}.

\subsubsection{Security analysis of modified COP}
Security analysis of modified COP has been performed against basic security features and integrated HBAT security features. To start with, we will focus on basic security features.\\

\textbf{Basic security features:} As discussed in original contribution \citep{acns}, basic security features mainly includes security against \textit{observation attack through brute force search} and \textit{password guessing attack}. Next we show how modified COP defends these attacks.\\

\textbf{(a) Security against observation attack through brute force search:}  In \cite{acns}, authors have determined the complexity of this attack as \large O(\normalsize n{$n+\ell-2 \choose \ell-1$}\large) \normalsize, where \textit{n} and $\ell$ denote the number of cells in the grid and length of the secret, respectively. As values of both these factors remain same in the proposed modification, thus modified COP provides same security standard as of the basic COP to resist this attack. \\

\textbf{(b) Security against password guessing attack:} In their analysis, in \cite{acns}, authors derive the probability of guessing the  correct secret as \large(\normalsize n \large  \normalsize {$n + \ell -2 \choose \ell-1$} \large)$^{-1}$ \normalsize. Therefore, it is easily understandable that due to no change in values of \textit{n} and $\ell$, proposed modification maintains the same security standard to resist this attack.\\

\textbf{Inherited HBAT security features:} Along with achieving flatness, integrated HBAT security feature includes security against DoS attack and MSV.\\

\textbf{(a) Security against DoS:} As modified COP utilizes $50\%$ of the total response set to create illusion in attacker's mind therefore, it can provide security against DoS with a probability of $1/2$.\\

\textbf{(b) Achieving flatness and security against MSV:} As modelling-syntax-approach may be used for creating the \textit{honeywords}, therefore though achieves complete flatness, but modified COP cannot provide adequate security against MSV.\\

\subsubsection{Usability analysis of modified COP} Usability analysis of modified COP is done against basic usability features and inherited HBAT usability features.\\

\textbf{Basic usability standard:} Modification of COP falls under the category of no \textit{CSLI} with no \textit{CLP} and no \textit{REI}. Thus, it provides the same usability standard as of the basic COP method.\\

\textbf{Inherited HBAT usability standard:} Proposed modification here neither imposes stress on user mind nor has any system interference. But probability of setting off a false alarm due to typing mistake of user becomes $4/9$ here which infers that this method is not much typo-safe. To increase typo-safety, system may ask a user to enter her response multiple (may be $2$) times so that probability of typing mistake becomes less.\\

\textbf{Note 7:} Proposed modification of COP does not make any impact on the basic security and usability standard of the basic COP scheme. From HBAT security point of view, it provides weak security against DoS which can be handled by adopting light security policy against this attack as shown in \citep{honeyword-erguler}. Modified COP here though cannot robustly handle the MSV crisis, but can achieve absolute flatness to lure the attacker properly. Form the HBAT usability aspect too, except typo-safety (may be increased though), modified COP meets with almost all the usability features.\\

\textbf{Takeaway:} Though a large number of methods fall under $\text{M}^\text{FODS}_\text{SOA}$ class, but with existing setup, they cannot provide any kind of security once $F_P$ gets compromised. Due to their design principle, migration to \textit{honeyword} based scheme is not immediate for these methods. By still using the concept of \textit{honeywords}, this paper deals with masking of plaintext password of the methods belonging to $\text{M}^\text{FODS}_\text{SOA}$ class . Influenced by the facts described in \textit{Note} $1$ and \textit{Note} $2$, we propose few principles in Section \ref{p-c-h}, by following which, any method belonging to $\text{M}^\text{FODS}_\text{SOA}$ class can store \textit{honeywords} in $F_P$. While \textit{Note} $3$ infers that proposed principles are sufficient for incorporating \textit{honeywords} to any method of the targeted class in theory, \textit{Note 4} to \textit{Note 7} suggest that HBAT framework can successfully be integrated with existing methods of $\text{M}^\text{FODS}_\text{SOA}$ class in practice too. 

Also, our proposal reveals that some methods (e.g., modified S3PAS) cannot choose a large value of $k$ as this may threat the usability standard. Though the large value of $k$ creates more confusion in attacker's mind, but we have shown that large value of $k$ makes it more likely to mount DoS attack. Hence, the value of $k$ needs to be chosen carefully to balance the security and usability aspects. As discussed previously, methods that provide weak security against DoS can adopt light security policy to resist this attack as advised in \citep{honeyword-erguler}. To mitigate typing error (especially for modified COP), a user may submit the derived response multiple time (may be twice) without allowing much degradation in usability standard.


\section{Conclusion}
\label{sec:conclusion}
Many fully observable defense mechanisms, identified as $\text{M}^\text{FODS}_\text{SOA}$, maintain password in plaintext format and migration to a \textit{honeyword} based scheme is not immediate for these methods. This makes any method of $\text{M}^\text{FODS}_\text{SOA}$ class vulnerable as the password can be easily obtained from the compromised password file. In this paper, our contribution overcomes this limitation of password maintenance mechanism by the $\text{M}^\text{FODS}_\text{SOA}$ category methods. 
We show how \textit{honeywords} can be used effectively to fill the existing security gap. To validate the practicality of our proposed solution, we have considered few well known shoulder surfing attack resilient defense mechanism which cannot afford to store \textit{honeywords} in their existing format. Modification of these schemes results in storing the \textit{honeywords} successfully. The modified schemes provide a satisfactory balance between usability and security features. Our study in this context also reveals an open problem regarding securing user's plaintext passwords from an adversary who can perform shoulder surfing attack even after obtaining the password file. Nevertheless, this paper shows use of \textit{honeywords} in a new direction and will motivate researchers for further usage of \textit{honeywords}.

\section*{References}

\end{document}